\pdfoutput=1

\documentclass{myarticle}
\usepackage{amsfonts,amssymb}
\usepackage{easybmat}
\usepackage{float}
\usepackage{amsmath}
\usepackage{lineno}
\usepackage{mathtools}
\usepackage{color,soul}
\usepackage{bookmark}
\usepackage{mathtools}
\usepackage{easybmat}
\usepackage{graphicx}
\usepackage{mathrsfs}
\usepackage{amssymb}
\usepackage{multirow}
\usepackage{array}
\usepackage{url}
\usepackage{color}
\usepackage[perpage,symbol]{footmisc}
\usepackage[position=t,singlelinecheck=off]{subfig}

\newcommand{\beginsupplement}{%
        \setcounter{table}{0}
        \renewcommand{\tablename}{\textbf{Supplementary Table}}
        \renewcommand{\thetable}{\textbf{\arabic{table}}}

        \setcounter{figure}{0}
        \renewcommand{\figurename}{\textbf{Supplementary Figure}}%
        \renewcommand{\thefigure}{\textbf{\arabic{figure}}}
     }

\setfnsymbol{wiley}

\title{A Tensor-Based Framework for Studying Eigenvector Multicentrality in Multilayer Networks}

\author{Mincheng Wu$^{1}$, Shibo He$^{1,}\footnote{To whom correspondence should be addressed. E-mail: s18he@zju.edu.cn, poor@princeton.edu.}$ , Yongtao Zhang$^1$, Jiming Chen$^1$, Youxian Sun$^1$, Yang-Yu Liu$^{2,3}$, Junshan Zhang$^4$ \& H. Vincent Poor$^{5,*}$}

\begin{document}

\maketitle

\begin{affiliations}
 \item State Key Laboratory of Industrial Control Technology, Zhejiang University, Hangzhou 310027, Zhejiang, China.
 \item Channing Division of Network Medicine, Brigham and Women's Hospital and Harvard Medical School, Boston, MA 02115.
 \item Center for Cancer Systems Biology, Dana-Farber Cancer Institute, Boston, MA 02115.
 \item School of Electrical, Computer and Energy Engineering, Arizona State University, Tempe, AZ 85287.
 \item Department of Electrical Engineering, Princeton University, Princeton, NJ 08544.
\end{affiliations}
\begin{center}
(\date{\today})
\end{center}

\begin{center}
\textbf{Abstract}
\end{center}
\begin{abstract}
Centrality is widely recognized as one of the most critical measures to provide insight in the structure and function of complex networks.
While various centrality measures have been proposed for single-layer networks, a general framework for studying centrality in multilayer networks (i.e., multicentrality) is still lacking.
In this study, a tensor-based framework is introduced to study eigenvector multicentrality, which enables the quantification of the impact of interlayer influence on multicentrality, providing a systematic way to describe how multicentrality propagates across different layers.
This framework can leverage prior knowledge about the interplay among layers to better characterize multicentrality for varying scenarios.
Two interesting cases are presented to illustrate how to model multilayer influence by choosing appropriate functions of interlayer influence and design algorithms to calculate eigenvector multicentrality.
This framework is applied to analyze several empirical multilayer networks, and the results corroborate that it can quantify the influence among layers and multicentrality of nodes effectively.

\end{abstract}

\clearpage
\tableofcontents
\clearpage

\section{Introduction}
Centrality quantifies  the importance of nodes in a graph, and has been widely studied to understand the structure and function of complex networks\cite{newman2003structure,boccaletti2006complex}.
For example, it can be utilized to  identify the most influential person in an online social network\cite{freeman1978centrality}, the most crucial artery in transport congestion\cite{sole2016congestion}, or the most important financial institution in the global economy\cite{battiston2012debtrank}.
Over 30 different  centrality measures (e.g., degree centrality, betweenness centrality, closeness centrality,  eigenvector centrality, and control centrality) have been examined  in the literature\cite{borgatti2006graph,lu2016vital,borgatti2005centrality,liu2012control}.
Among these, eigenvector centrality, defined as the leading eigenvector of the adjacency matrix of a graph, has received  increasing attention\cite{bonacich2007some,fraschini2015eeg}.
It is worth noting that PageRank, a variant of  eigenvector centrality,
is the primary algorithm used in Google's search engine\cite{brin2012reprint,langville2008google}.

Notably, most previous studies have focused on eigenvector centrality in a single-layer network, in which all nodes/links are assumed to be of the same type (centrality-homogeneous).
As revealed recently\cite{buldyrev2010catastrophic,kivela2014multilayer,boccaletti2014structure,nicosia2015measuring,pilosof2017multilayer,gao2017complex}, many practical complex systems, ranging from the Internet to airline networks, have multiple types  of nodes and/or  links between nodes.
Multilayer networks, which consist of multiple layers of nodes with intra- and interlayer links, can be used to model such  complex systems.
Figure \ref{fig.examples} shows two examples of multilayer networks (see Supplementary Fig. 1 for more examples).
Simply aggregating a multilayer network into a single-layer one would obviously lead to a miscalculation of centrality.
Recent work on eigenvector-like centrality in multilayer networks either  assigned constant weights to predetermine interlayer influence (which can be regarded as the gain or loss of the interplay strength between two layers\cite{sola2013eigenvector}), or focused on a special case of multilayer networks, i.e., the so-called \textit{multiplex networks} (where all layers share the same set of nodes, and  interlayer links only exist between counterpart nodes)\cite{sole2014centrality,zhou2007co,de2015ranking,iacovacci2016functional}.
It is of significant interest to develop a framework for studying eigenvector-like centrality in general multilayer networks, hereafter  referred to as eigenvector multicentrality.
In this study, we introduce a tensor-based framework that enables the quantification of the relationship between interlayer influence and eigenvector multicentrality.
It is challenging to compute eigenvector multicentrality of nodes in such a framework since  interlayer influence  and eigenvector multicentrality are interdependent.
We prove the existence and uniqueness of eigenvector multicentrality for given appropriate forms of interlayer influence.
We also design efficient algorithms to calculate it for two interesting scenarios.
This framework offers a novel approach for  modeling and quantifying the interlayer interactions in multilayer networks, providing a systematic way of characterizing eigenvector multicentrality.
Experimental results based on several real-world multilayer networks corroborate our analytical results.

\clearpage
\section{Results}
\subsection{A tensor-based framework for studying eigenvector multicentrality.}

In the calculation of eigenvector centrality of nodes in a single-layer network, a directed link to a node can be viewed as  a vote of support.
Each node fairly propagates its entire centrality score to its neighbors recursively. 
The eigenvector centrality of a node is defined as the scores that it gathers  from its neighbors after appropriate normalization in the steady state.
Formally, the vector consisting of the eigenvector centrality of all nodes is defined as the leading left eigenvector of the adjacency matrix associated with the single-layer network.
We generalize this definition to multilayer networks by taking into account interlayer influence among layers.
Specifically, a multilayer network is modeled as $\mathcal{M}=(\mathcal{L},\mathcal{E})$, where $\mathcal{L}=\{L_\alpha;\alpha=1,2,\cdots, K\}$ is a collection of graphs $L_\alpha=(V_\alpha, E_\alpha)$ representing layers in $\mathcal{M}$;  $V_\alpha=\{v_{1,\alpha},v_{2,\alpha},\cdots v_{n_\alpha,\alpha}\}$ is the set of nodes, where $n_\alpha$ denotes the number of nodes, and $E_\alpha$ is the set of intralayer links in layer $\alpha$; $\mathcal{E}=\{E_{\alpha\beta}\subseteq V_\alpha\times V_\beta; \alpha,\beta =1,2,\cdots,K (\alpha\ne \beta) \}$ contains the interlayer links in $\mathcal{M}$\cite{boccaletti2014structure}.
To avoid confusion, we use Latin letters $\{i,j,\cdots\}$ to indicate nodes and Greek letters $\{\alpha,\beta,\cdots\}$ to indicate layers.
Tensors provide a general mathematical tool to describe high-dimensional objects\cite{de2013mathematical,beigi2015on}, and notably, tensors have been employed to study multilayer networks\cite{valdano2015analytical,de2016degree,de2017disease}.
For example, a fourth-order tensor $M_{j\beta}^{i\alpha}$, called the adjacency tensor, is used to encode a directed, weighted link from node $i$ in layer $\alpha$ to node $j$ in layer $\beta$ (see Supplementary Text I for further details on tensorial representations).
We further introduce the influence tensor $W^\alpha_{\beta}$, which is a second-order tensor measuring the interlayer influence from layer $\alpha$ to layer $\beta$.
In our framework, the influence tensor $W$ may be treated as a constant tensor when quantitative knowledge is available.
The interaction tensor is defined as $H^{i\alpha}_{j\beta}=W^\alpha_{\beta} M_{j\beta}^{i\alpha}$, encoding the interaction from node $i$ in layer $\alpha$ to node $j$ in layer $\beta$.
Intuitively, when the influence from  layer $\alpha$ to layer $\beta$ is greater than one, the centrality scores propagating along the links from layer $\alpha$ to  layer $\beta$ will be magnified,  and vice versa.

The second-order tensor $\Phi_{i\alpha}$ is defined as the solution to the following tensorial equation:
\begin{equation}\label{eq.frame}
\begin{split}
&H_{j\beta}^{i\alpha}\Phi_{i\alpha}=\lambda_{\beta}\Phi_{j\beta},
\end{split}
\end{equation}
where $\lambda_{\beta}$ is a coefficient related to layer $\beta$,
and the Einstein notation\cite{aahlander2002einstein} is adopted here (see Supplementary Text I for further details).
Because $\Phi$ is a eigenvector-like centrality, it will hereafter be referred to  as eigenvector multicentrality, and $\Phi_{i\alpha}$ represents the eigenvector multicentrality score of node $i$ in layer $\alpha$.
In  the calculation of  eigenvector multicentrality, after each node propagates  its entire multicentrality score to neighbors, the scores from layer $\alpha$ to layer $\beta$ will be multiplied by the influence coefficient $W^\alpha_{\beta}$. 
Hence, we can obtain the eigenvector multicentrality via appropriate normalization in the steady state.
Notice that the normalizing coefficient $\lambda_{\beta}$ may be different for different layers in a multilayer network. 
This differs from the eigenvector centrality in a single-layer network, where all nodes share a common normalizing coefficient $\lambda_{1}$ (namely, the leading eigenvalue of the adjacency matrix).

Many existing models can be incorporated in our eigenvector multicentrality framework  by choosing the influence tensor $W$ appropriately.
For instance, in an author-document heterogeneous network, unweighted interlayer links connect documents to their authors.
A directed unweighted intralayer link exists between two documents if one document refers to the other and the undirected weighted intralayer link between two authors represents their social tie.
The multicentrality in such an author-document network was defined as the leading eigenvector of a stochastic matrix, which encodes the probabilities that a random surfer moves along intra- and interlayer links  in a combined random walk process\cite{zhou2007co}. Clearly, this model can be incorporated in our framework by setting the intra- and interlayer influence weights to predetermined constants.
Further, a definition of multicentrality in multiplex networks has been proposed in Ref. 20, considering the influence from counterpart nodes in other layers by importing the influence matrix $Q=(q_{\alpha\beta})\in \mathbb{R}^{K\times K}$. The influence matrix is non-negative, and $q_{\alpha\beta}$ measures the influence of layer $\beta$ on layer $\alpha$.
One can calculate eigenvector-like multicentrality of a multiplex network once $Q$ has been obtained. 
Observe that in Ref. 20 the influence matrix $Q$ is predetermined and given as in a lower-dimensional form (i.e., a matrix), whereas in our framework the influence tensor $W$ depends on the eigenvector multicentrality and hence they are interdependent.
Moreover, interconnected multilayer networks have been proposed to predict diffusive and congestion processes\cite{de2015ranking}, where the undirected unweighted interlayer links connect nodes to their counterparts in other layers. 
Eigenvector centrality in these networks  is a special case of our framework where the interlayer influence is equal to one.

\subsection{Leveraging prior knowledge about interlayer interactions.}
Quantifying the influence tensor $W$ is important to calculate multicentrality in our framework.
As expected, the influence tensor $W$ is typically a function of the adjacency tensor $M$ and the multicentrality $\Phi$, rather than being a constant.
In practice, precisely predetermining the influence tensor is often infeasible.
One advantage of our general framework lies in the leveraging of prior knowledge about the influence tensor $W$ in diverse applications and the  calculation of multicentrality even when $W$ and $\Phi$ are interdependent.

Consider a typical scenario in which all nodes are centrality-homogeneous and layers are heterogeneous in a multilayer network.
In such a scenario, the multicentrality scores of all nodes are comparable, and can be represented by a vector $C\in \mathbb{R}^N$, where $N$ is the number of all nodes. We call the eigenvector multicentrality in such a scenario  global multicentrality. 
We normalize the multicentrality such that the vector $C$ is defined over an $N$-dimensional simplex.
For example, in a multilayer network consisting of web pages on different subjects (see the empirical results for further details), we may need to compare the multicentrality scores of two web pages on different subjects.
Clearly, the multicentrality score propagates differently along interlayer links and along intralayer links, owing to differences in the popularity of different subjects. 
Here, we assume that the importance of layer $\alpha$ is a function of the multicentrality of all nodes in layer $\alpha$, denoted by $f(\Phi_{:\alpha})$, where the colon $``:"$ indicates all elements of a given dimension\cite{kolda2009tensor}. 
Notably, the function $f(\cdot)$ describing the layer importance is application-dependent.
The function $f$ could be, for example, the $L^1$-norm $f({\Phi_{:\alpha}})=\|\Phi_{:\alpha}\|_1$, which means the aggregated multicentrality scores of nodes in layer $\alpha$, or it could be $f({\Phi_{:\alpha}})=\|\Phi_{:\alpha}\|_1/n_\alpha$, which denotes the average multicentrality score over nodes in layer $\alpha$.

There are a variety of ways to define the influence tensor. In this article, we define
\begin{equation}\label{eq.influence}
\begin{split}
W^\alpha_{\beta}=f(\Phi_{:\alpha})/f(\Phi_{:\beta}).
\end{split}
\end{equation}
That is to say, the interlayer influence between two layers depends on their relative layer importance.
It is clear that there is  no gain or loss for links between nodes in the same layer,  because the interlayer influence $W^\alpha_{\alpha} (\alpha=1,2,\cdots,K)$ is equal to one.
For  links from a node in a more important layer, there is a gain in the multicentrality, and vice versa.
Then the interaction tensor can be written as  $H^{i\alpha}_{j\beta}=\frac{f(\Phi_{:\alpha})}{f(\Phi_{:\beta})}\cdot M_{j\beta}^{i\alpha}$. Further, we prove that $\lambda_\alpha=\lambda_1, \forall \alpha\in\{1,2,\cdots, K\}$, and Eq. \ref{eq.frame} reduces to
\begin{equation}\label{eq.main}
\begin{split}
H_{j\beta}^{i\alpha}\Phi_{i\alpha}=\lambda_1\Phi_{j\beta},
\end{split}
\end{equation}
where $\lambda_1$ is the leading eigenvalue of the interaction tensor $H$, and is irrelevant to $\beta$ in this scenario.
Notably, $\lambda_1$ is also the leading eigenvalue of the adjacency tensor $M$, indicating that the interaction tensor $H$ maintains the leading eigenvalue of the adjacency tensor $M$ (see Supplementary Text V for further details).
From Eq. \ref{eq.main}, we can see how multicentrality score propagates in a multilayer network.
Considering that a node $i$ in layer $\alpha$ links to another node $j$ in layer $\beta$, node $i$ will propagate its multicentrality score to node $j$ scaled by an influence coefficient $W^\alpha_{\beta}$, which could be a gain ($W^\alpha_{\beta}>1$), a loss ($W^\alpha_{\beta}<1$), or even ($W^\alpha_{\beta}=1$), and the multicentrality $\Phi_{j\beta}$ comprises the scores that node $j$ in layer $\beta$ gathers in the steady state.
The existence and uniqueness of the multicentrality $\Phi$ are proved in Supplementary Text V. 
For a given function $f$, we can calculate the global multicentrality using the compressed power iteration method introduced in the Materials and Methods section.

We consider another interesting scenario in which nodes in different layers are  heterogeneous, and thus are not comparable. For example, in a heterogeneous network consisting of authors and papers, it is not meaningful to compare the multicentrality of an author with that of a paper.
Because the multicentrality of nodes in different layers may have varying implications, we can only calculate the local multicentrality of nodes in each layer while taking into account the interlayer influence, where local multicentrality means that the nodes in each (local) layer are centrality-homogeneous.
In such a scenario, the multicentrality score of a node cannot simply propagate along interlayer links to other layers.
We measure the local multicentrality of nodes in each layer by defining the influence tensor  in the framework as
\begin{equation}\label{eq.local}
\begin{split}
W^\alpha_{\beta}=\frac{\sum_{i,j=1}^NM^{j\beta}_{i\alpha}\Phi_{j\beta}}{\sum_{i,j=1}^NM^{i\alpha}_{j\beta}\Phi_{i\alpha}}.
\end{split}
\end{equation}
Notice that the denominator $\sum_{i,j=1}^NM^{i\alpha}_{j\beta}\Phi_{i\alpha}$ in $W^\alpha_{\beta}$ is a normalizing constant, which is the sum of scores propagating along interlayer links from layer $\alpha$ to layer $\beta$. 
Moreover, the sum of interactions from nodes in layer $\alpha$ to nodes in layer $\beta$ is given by the numerator $\sum_{i,j=1}^NM^{j\beta}_{i\alpha}\Phi_{j\beta}$, which is the sum of scores propagating from layer $\beta$ to layer $\alpha$.
Therefore, by defining the influence tensor in Eq. \ref{eq.local}, we assume that the score flow going out of one layer will be returned to the layer.
In such a way, we can calculate the local multicentrality of nodes in each layer independently while the interlayer influence is taken into account.
The detailed proofs of the existence and uniqueness of $\Phi$ in local multicentrality are provided in Supplementary Text VI.
We can also calculate the local multicentrality $\Phi$ numerically using the compressed power iteration method.

Because the prior knowledge is network-specific and application-dependent, we present two interesting scenarios, for global multicentrality and local multicentrality respectively, to illustrate how to leverage prior knowledge to find $W$ and compute the multicentrality of nodes.
The PageRank algorithm has been now widely utilized in social, transportation, biology and information network analysis for link prediction, recommendation, etc.\cite{gleich2015pagerank}.
Note that PageRank centrality is a variant of eigenvector centrality.
In PageRank centrality, each node distributes its PageRank score to its neighbors along outgoing links on an equal footing, and a node's PageRank score is defined as the sum of scores that it gathers from its neighbors in the steady state.
Our eigenvector multicentrality framework can be easily carried over to characterize PageRank multicentrality (see Supplementary Text IV for further details about PageRank multicentrality).

\subsection{Multicentrality in empirical networks.}

We first  consider a dataset from Wikipedia consisting of 4,604 web pages (see the Materials and Methods section for further details about this dataset and how it has been obtained).
The web pages are divided by Wikipedia into 15 subjects, including \textit{art, business studies, citizenship, countries, design and technology, everyday life, geography, history, IT, Language and literature, mathematics, music, people, religion} and \textit{science}, and one web page belongs solely to one subject.
We build a multilayer network by placing web pages of the same subject in the same layer, and establishing a directed link between two web pages if there is a hyperlink between them (a subnetwork is shown in Fig. 1A).
For comparison, we also build a single-layer network by aggregating all web pages in all layers (see Supplementary Fig. 4 for further details).
We select 4,128 web pages from these, in order to guarantee that the network is connected and all nodes have at least one out-degree and one in-degree.

We measure the PageRank multicentrality of web pages in the constructed multilayer network.
Moreover, because web pages in different subjects are centrality-homogeneous, global PageRank multicentrality is used.
We consider three common forms for layer importance $f(\cdot)$:
$f_1(\Phi_{:\alpha})=\ln(1+N\cdot|\Phi_{:\alpha}|_1/n_\alpha)$, $f_2(\Phi_{:\alpha})=|\Phi_{:\alpha}|_\infty$ and $f_3(\Phi_{:\alpha})=|\Phi_{:\alpha}|_1/n_\alpha$.
Table 1 shows the results of global PageRank multicentrality ($f=f_1$) in the Wikipedia multilayer network and PageRank centrality in the aggregated Wikipedia network, where the digits in parenthesis indicate the differences between these two rankings (see Supplementary Tables 1-3 for further details).

Note that the entry ``United States'' has the largest multicentrality score, because  it has the largest number of incoming links. Furthermore, the  web pages linked to it have high multicentrality scores.
The entry ``Europe'' in the layer ``Geography'' has many enhanced links from  nodes in the layer ``Countries'', because the average PageRank multicentrality score of nodes in ``Countries'' is higher than that in ``Geography''.
``France'' has  large numbers of  incoming links from entries in the layers ``Art'', ``Music'' and other subjects; scores along these incoming links, however, will be diminished since ``Art'' and ``Music'' are less important than ``Countries''.
Note that the entry ``Television'' is significantly promoted, because the layer importance of ``Design and Technology'' is relatively low. 
Global multicentrality can effectively quantify the influence between layers even when we have limited prior knowledge, rather than aggregating all nodes without considering interlayer influence, as shown in many previous methods.

Eigenvector multicentrality is an effective predictor for searching for the most important node in multilayer networks, which is also validated by the results obtained from the page views (PVs) in Wikispeedia.
Wikispeedia is a human-computation game\cite{west2009wikispeedia}, in which users are requested to navigate from a given web page to a  target one by only clicking on Wikipedia links. 
We collect all completed navigation paths and obtain the PVs of each web page from Wikispeedia.
For all entries, we calculate their PageRank multicentrality scores, PageRank centrality and degree centrality scores, and compare them to their PVs  in Wikispeedia (see Supplementary Tables 4-8 for further details). 
The results are shown in Figs. \ref{fig.wiki} and \ref{fig.scatter}.
We also list the average PageRank multicentrality score of each layer (subjects) in Wikipedia (see Supplementary Tables 9-11 for further details).
The Spearman rank correlation  coefficients show that PageRank multicentrality outperforms PageRank centrality and degree centrality in the aggregated network.

\begin{figure*}
\centering
\includegraphics{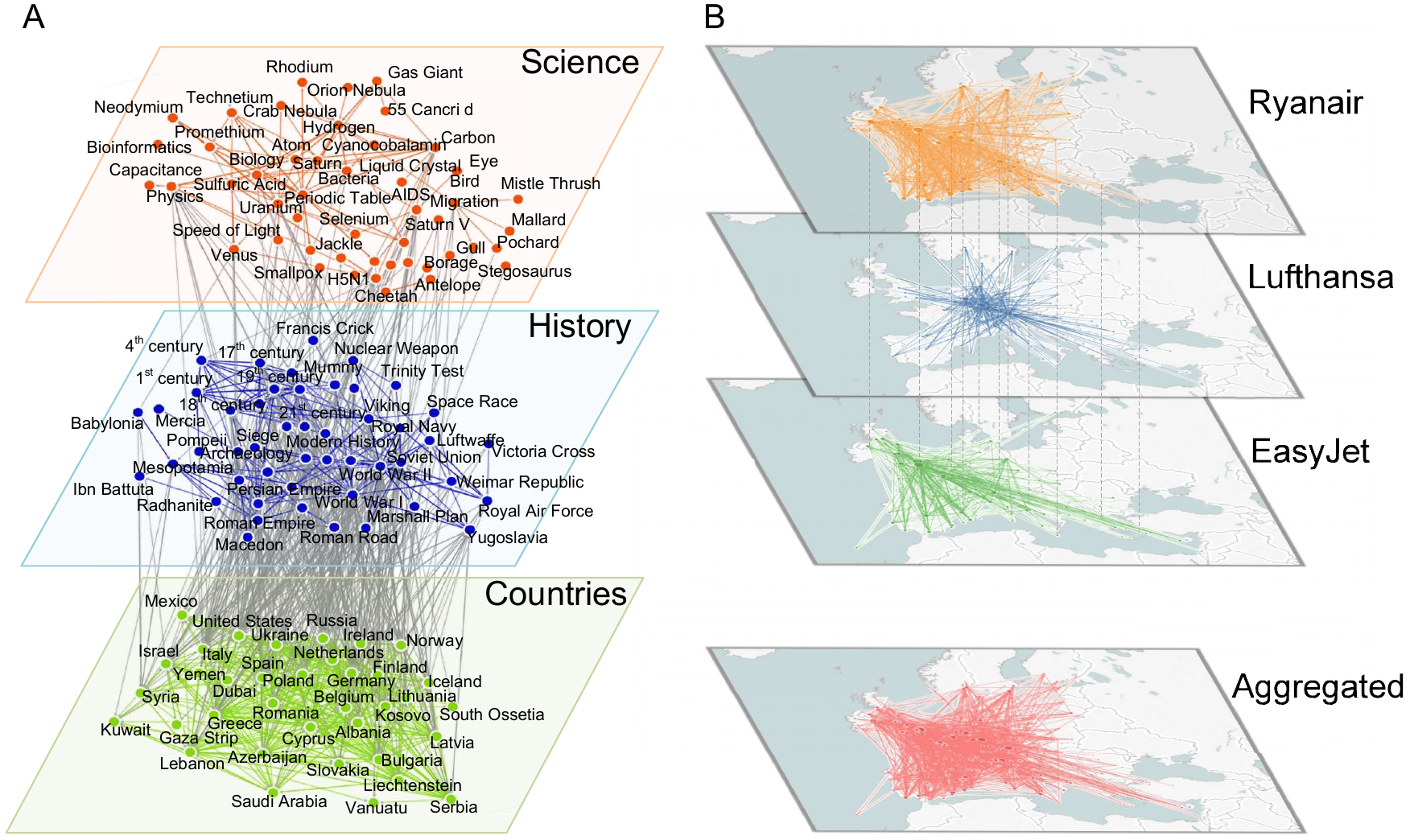}
\caption{Examples of multilayer networks. (\textbf{A}) A network of web pages in Wikipedia  can be considered as  a multilayer network. The different layers represent different subjects, and  nodes denote words (or terms) connected by hyperlinks. Colorized links are intralayer links while  gray ones are interlayer links.
(\textbf{B})  A European airline network with three layers can be modeled as a multiplex network, which contains the same set of nodes in all layers. Intralayer links in one layer represent flight routes operated by  one airline  and interlayer links only exist between the same nodes (airports)  in different layers. Only a portion of interlayer links are shown in order not to complicate the figure. A single-layer network obtained by simply aggregating all airports and flight routes is also shown at the bottom. The geographic data is provided by OpenStreetMap.}\label{fig.examples}
\end{figure*}

\begin{figure*}
\centering
\includegraphics{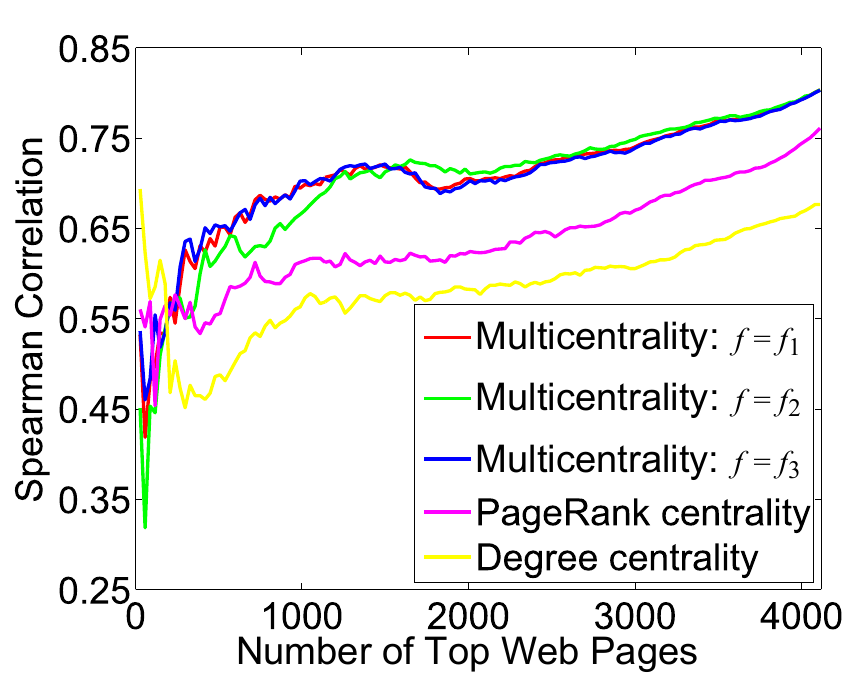}
\caption{PageRank multicentrality, PageRank centrality and degree centrality in the Wikipedia multilayer network. The line chart shows the Spearman correlation coefficients for different number of top web pages, where the horizontal axis denotes the number of top nodes that we select,  and the vertical  axis represents the corresponding Spearman correlation coefficients. When the number of nodes is smaller than 200 (5\% of the total nodes), the degree centrality has better performance. However, when more nodes are involved, the multicentrality performs better. In particular, the Spearman correlation coefficients reach  0.80 for all nodes under the three proposed multicentrality measures.}\label{fig.wiki}
\end{figure*}

\begin{figure*}
\centering
\includegraphics{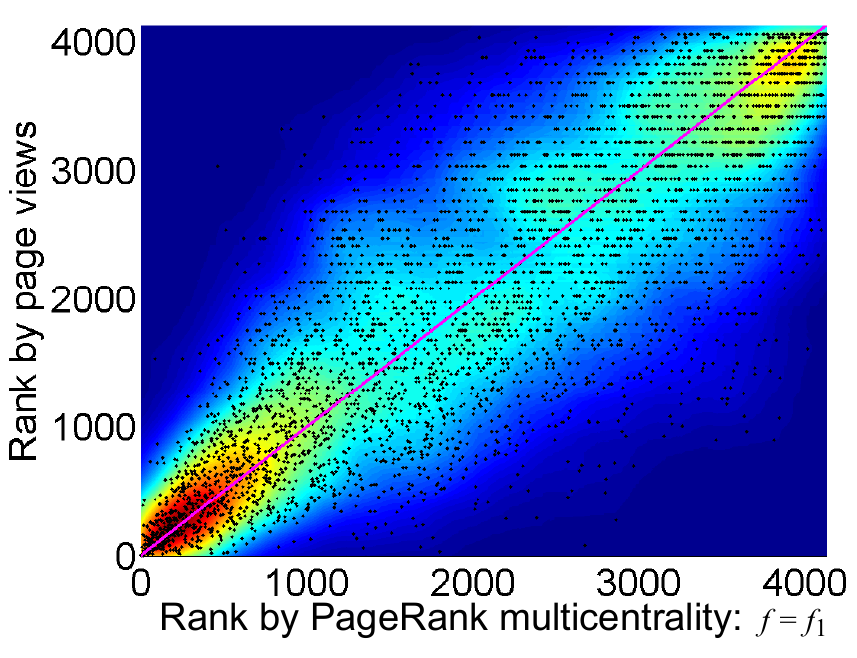}
\caption{Scatter diagram of the ranks of the entries by multicentrality and by PVs. The diagram shows the PageRank multicentrality of all nodes when $f=f_1$ (see Supplementary Fig. 5 for additional diagrams), where the horizontal axis denotes the rank of the entries by  multicentrality and the vertical axis represents the rank of the entries by PVs in Wikispeedia. The Spearman correlation coefficient between the two rankings is 0.80. }\label{fig.scatter}
\end{figure*}

\begin{figure*}
\begin{center}
\includegraphics{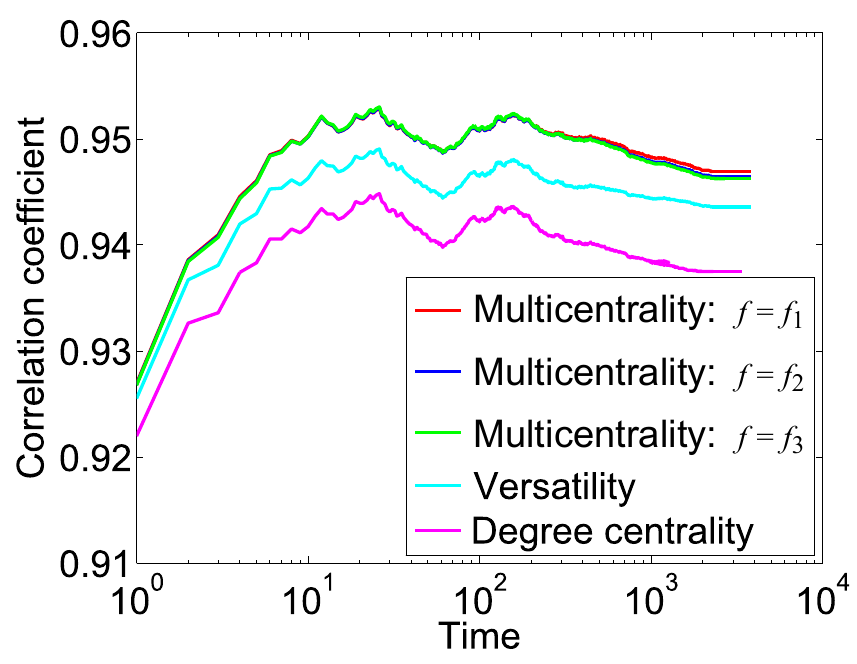}
\end{center}
\caption{Comparison of PageRank multicentrality, PageRank versatility and degree centrality for the European airline network. The Spearman correlation coefficients in the dynamical process are plotted.   When $t\geq3,000$, these five curves of correlation coefficients tend to be stable, and there is a clear gap between the multicentrality measure and other measures.}\label{fig.euair}
\end{figure*}

Next, we consider a transportation network consisting of airports and air routes between them. We first consider 450 airports in Europe\cite{cardillo2013emergence} (see the Materials and Methods section for details about this dataset and how it has been obtained), and we focus on three main airlines. 
We then build a multiplex network with three layers (airlines) and  450 nodes (airports) in each layer as shown in Fig. \ref{fig.examples}B, where the dotted lines are interlayer links between airports and their counterparts in other layers.
We measure the global PageRank multicentrality in the multiplex network using three forms of layer importance: $f_1(\Phi_{:\alpha})=e^{|\Phi_{:\alpha}|_1/n_\alpha}-1$, $f_2(\Phi_{:\alpha})=|\Phi_{:\alpha}|_1/n_\alpha$
and $f_3(\Phi_{:\alpha})=\ln(1+N\cdot|\Phi_{:\alpha}|_1/n_\alpha)$.
Then, the PageRank multicentrality score of each airport is obtained by assembling the multicentrality scores of all its counterparts in all layers.
For comparison, we also build a single-layer network, called the aggregated network, by combining the same airports in the three layers.
We further introduce the versatility, a good predictor for diffusive and congestion processes in multilayer networks\cite{de2015ranking}, which is a special case of our framework when setting  all components  of the influence tensor $W$ to one.
We focus on the coverage $\rho(t)$, a suitable proxy for the exploration efficiency of the network\cite{de2014navigability}, defined as the average fraction of distinct nodes being visited up to time $t$ regardless of the layer,  assuming that a walker starts from a certain node in the network
(see the Materials and Methods section for further details about the coverage).
We investigate whether  multicentrality helps understand the role that a node plays in dynamical scenarios.
To this end, we compute the Spearman correlation coefficient between the ranking of the airports by multicentrality and that by the coverage at time $t$ of a hypothetical epidemic spreading process that starts from a certain airport.
For comparison, we also compute baselines such as the rankings by versatility and PageRank centrality in the aggregated network (see Supplementary Tables 12-14 for further details).
We calculate the Spearman correlation coefficients for these five methods at each time step, and the results are shown in Fig. \ref{fig.euair}, where the time ranges from $t=1$ to $t=4,000$.
It is shown that the three multicentrality measures achieve higher accuracy (their  correlation coefficients exceed $0.947$) in the steady state ($t\geq3,000$).
We then perform a similar analysis on an airline network from the U.S., which contains the airlines flying from the U.S. on Jan 3, 2008 (with data provided by the American Statistical Association Sections on Statistical Computing, \url{http://stat-computing.org/}). 
We build a multiplex network with 20 layers and  284 nodes in each layer, and a similar conclusion can be drawn (see Supplementary Table 15 for further details).

Another real-world example we consider is a social network, where we apply our framework to an  e-mail network constructed from a large European research institution with 1,005 nodes (individuals) and 42 layers (departments), on which we consider an epidemic spreading process. 
The simulation results indicate that the nodes with higher eigenvector multicentrality play a more important role in the  epidemic spreading process (see Supplementary Text VII for details).

\section{Discussion}
As shown in recent work on eigenvector centrality (and its variants)\cite{taylor2017eigenvector,aguirre2013successful,iyer2013attack}, it is of significant interest to build a framework for studying eigenvector-like centrality in multilayer networks.
The existing studies, however, assumed empirical influence coefficients or relied on specific types of multilayer networks.
Here we develop a general framework for studying eigenvector multicentrality in multilayer networks, which enables the quantification of the impact of interlayer influence on eigenvector multicentrality, providing an analytical tool to describe how eigenvector multicentrality propagates among different layers.
Further, this framework can easily leverage prior knowledge about the interplay among layers to characterize eigenvector multicentrality for varying scenarios.
As the interlayer influence and multicentrality of nodes are interdependent, they are jointly solved using a compressed power iteration method.
Furthermore, we formulate and analyze PageRank multicentrality for practical applications within the proposed framework.
We also perform theoretical analyses to prove the existence and uniqueness of the solutions in Supplementary Text V-VI, which allows us to calculate global and local multicentrality in any strongly connected multilayer networks. 
For an arbitrary multilayer network, we treat a dead end (a node with no outgoing links) the same as if it had outgoing links to all nodes, and introduce a damping factor to guarantee the existence and uniqueness of the solution.
The results from  empirical networks demonstrate that our general framework can effectively quantify the interlayer influence, and eigenvector multicentrality is a good measure to identify important nodes from both structural and dynamical perspectives.
Thus,  multicentrality aids in understanding and predicting the behaviors of dynamic processes by leveraging network structure, and describes the structure-function relationship of multilayer networks well.

We believe that the concept of multicentrality has the potential to enable a deep understanding of the structure and function of multilayer networks.
Because the real-world scenarios of multilayer networks vary, a key step is to find appropriate forms of interlayer influence for theoretical analyses. Here we  consider two interesting scenarios, for global multicentrality and local multicentrality respectively.
We believe that our proposed tensor-based framework can be applied to  more empirical networks in various  scenarios, including social networks, transportation networks, biological networks, etc.

\section{Materials and Methods}
\subsection*{Numerical solution.}
The crux of the proposed framework is to solve the tensorial equation
\begin{equation}
H_{j\beta}^{i\alpha}\Phi_{i\alpha}=\lambda_{\beta}\Phi_{j\beta},
\end{equation}
where the solution $\Phi$ is  the multicentrality tensor.
To obtain the numerical solution, we first flatten the adjacency tensor $M$ into a matrix, i.e., we represent the fourth-order tensor $M\in\mathbb{R}^{N\times N\times K\times K}$ as a matrix $\overline{M}\in \mathbb{R}^{NK\times NK}$, where $\overline{M}$ denotes the lower-dimensional form of the tensor $M$.
Then we vectorize the second-order tensor $\Phi\in\mathbb{R}^{N\times K}$ into a supravector $\overline{\Phi}\in \mathbb{R}^{NK}$ and denote by $\overline{W}$ the matrix form of the influence tensor $W$ (see Supplementary Text II for further details of  tensor decomposition). Further, we denote by $\Lambda=[\lambda_{1},\lambda_{2},\cdots,\lambda_{K}]^\top\in\mathbb{R}^{K}$  a vector encoding the  normalizing coefficient in each layer.
Thus, we obtain the matrix equation
\begin{equation}\label{matrixequation}
(\overline{W}\odot\overline{M})^\top\cdot\overline{\Phi}=\Lambda\odot\overline{\Phi},
\end{equation}
where $\odot$ denotes the Khatri-Rao product and $\overline{W}=\overline{W}(\overline{M},\overline{\Phi})$ is  a function of the matrix $\overline{M}$ and the multicentrality $\overline{\Phi}$.

For the numerical solution, we propose a compressed power iteration method, whose iteration scheme is as follows:
\begin{equation}
\overline{\Phi}^{(k+1)}=\overline{\Phi}^{(k)}+D^{(k)}[(\overline{W}\odot\overline{M})^{\top(k)}\odot{\Omega^{(k)}}-E_N]\overline{\Phi}^{(k)},
\end{equation}
where $\Omega^{(k)}=[\omega^{(k)}_1,\omega^{(k)}_2,\cdots,\omega^{(k)}_K]\in \mathbb{R}^K$ is the normalizing vector. Denoting $B^{(k)}=(\overline{W}\odot\overline{M})^{\top(k)}\odot{\Omega^{(k)}}-E_N$, we can write the iteration scheme as
\begin{equation}\label{iteration}
\begin{split}
\overline{\Phi}^{(k+1)}=\overline{\Phi}^{(k)}+D^{(k)}B^{(k)}\overline{\Phi}^{(k)},
\end{split}
\end{equation}
where $D^{(k)}\in \mathbb{R}^{N\times N}$ is a diagonal matrix related to $B^{(k)}$, and $D^{(k)}$ compresses the induced infinity norm of the matrix $B^{(k)}$ such that the $L^1$-norm of each row in $D^{(k)}B^{(k)}$ is strictly less than one.

Specifically, we let $\omega^{(k)}_\gamma=\|\overline{\Phi}_{\gamma}^{(k)}\|_1^{-1}$. With regard to the global multicentrality, the vector $\Lambda$ contains equivalent elements, i.e., $\Lambda=[\lambda_{1},\lambda_{1},\cdots,\lambda_{1}]^\top$. Thus we have $\Omega^{(k)}=[\omega^{(k)}_1,\omega^{(k)}_1,\cdots,\omega^{(k)}_1]^\top$, where $\omega^{(k)}_1=\|\overline{\Phi}^{(k)}\|_1^{-1}$.
Further, we can specify the diagonal matrix $D^{(k)}$ to compress the induced infinity norm of matrix $B^{(k)}$, where  $D^{(k)}$ is not unique in practice. For example, we could take  $D^{(k)}=({\rm diag}\{(B^{(k)}+E_N)\cdot\textbf{1}_{N\times 1}\})^{-1}$.
Then for each $\overline{\Phi}^{(k)}>0$ (i.e., all the components in $\overline{\Phi}^{(k)}$ are positive), the matrix $[E_N+D^{(k)}B^{(k)}]$ is strictly diagonal dominant with positive elements.
Hence, the iterations in Eq. \ref{iteration} converges to a unique solution\cite{Ruggiero1990an,berman1994nonnegative} and this solution satisfies the matrix Eq. \ref{matrixequation} (see Supplementary Text VIII for further details).

\subsection*{Multilayer network of  Wikipedia.}
The Wikipedia dataset  contains 4,604 entries and 119,882 hyperlinks\cite{west2012human} (data provided by the Stanford Network Analysis Project, \url{http://snap.stanford.edu/index.html}).
To ensure connectivity, we select the entries that have at least one out-degree and one in-degree. Then 4,128 entries and 113,441 links are obtained and these entries are divided into 15 subjects by Wikipedia.
Resultantly, we can build a multilayer network with $N=4,128$ nodes and $K=15$ layers, where each layer contains the entries of a single subject. The adjacency tensor $M \in \mathbb{R}^{N\times N\times K\times K}$ encodes the directed and unweighted links of the multilayer network and the influence tensor $W \in \mathbb{R}^{K\times K}$ encodes the interlayer influence between any two layers.
We measure the interlayer influence
\begin{equation}\label{method.it}
\begin{split}
W^\alpha_{\beta}=f(\Phi_{:\alpha})/f(\Phi_{:\beta}),
\end{split}
\end{equation}
where $f(\Phi_{:\alpha})$ indicates the layer importance of layer $\alpha$.
Here, we consider three forms of layer importance:
$f_1(\Phi_{:\alpha})=\ln(1+N\cdot|\Phi_{:\alpha}|_1/n_\alpha)$, $f_2(\Phi_{:\alpha})=|\Phi_{:\alpha}|_\infty$ and $f_3(\Phi_{:\alpha})=|\Phi_{:\alpha}|_1/n_\alpha$.
Further, we obtain the interaction tensor $H \in \mathbb{R}^{N\times N\times K\times K}$ as $H^{i\alpha}_{j\beta}=W^\alpha_{\beta} M_{j\beta}^{i\alpha}$.
Following the construction of the interaction tensor $H$, we then solve the tensorial equation
\begin{equation}\label{method.main}
\begin{split}
H_{j\beta}^{i\alpha}\Phi_{i\alpha}=\lambda_1\Phi_{j\beta}
\end{split}
\end{equation}
using the compressed iteration method.
Finally, the multicentrality tensor $\Phi$ is in the space $\mathbb{R}^{N\times K}$, and  $\Phi_{i\alpha}$ represents the multicentrality of  node $i$ in  layer $\alpha$.

\subsection*{Multilayer network of the European airlines.}
The European airline network contains  450 airports in Europe and the air routes for 37 airlines (see Ref. 34 for more details about this dataset).
For each airline, we can build a network with $N=450$ nodes and a set of links representing routes between  airports.
We select those airlines with the number of air routes greater than $N/2$, such that the average degree for each node  in the constructed network is at least one. 
In this way, we obtain three main airlines: \textit{Ryanair}, \textit{Lufthansa} and \textit{Easyjet}.
We thus have a multiplex network with $K=3$ layers and 450 nodes in each layer.
Then, we interconnect  the same airport across layers, obtaining a three-layer multiplex network.
The adjacency tensor $M$ is in the space $\mathbb{R}^{N\times N\times K\times K}$.
In the context of the  multiplex network, $M_{j\alpha}^{i\alpha}$ encodes the undirected and unweighted intralayer links in layer $\alpha$, while $M_{j\beta}^{j\alpha}=1$ encodes the undirected and unweighted interlayer link for node $j$ between layer $\alpha$ and layer $\beta$.
For the interlayer influence, we again consider three forms of the layer importance:
$f_1(\Phi_{:\alpha})=e^{|\Phi_{:\alpha}|_1/n_\alpha}-1$, $f_2(\Phi_{:\alpha})=|\Phi_{:\alpha}|_1/n_\alpha$
and $f_3(\Phi_{:\alpha})=\ln(1+N\cdot|\Phi_{:\alpha}|_1/n_\alpha)$. After obtaining the influence tensor $W$ via  Eq. \ref{eq.influence}, we have the interaction tensor $H^{i\alpha}_{j\beta}=W^\alpha_{\beta} M_{j\beta}^{i\alpha}$ in the space $\mathbb{R}^{N\times N\times K\times K}$.
Finally, we solve the tensorial Eq. \ref{eq.main} using the compressed iteration method and obtain the multicentrality tensor $\Phi_{i\alpha}$.

With respect to the coverage $\rho(t)$, we have
\begin{equation}\label{method.coverage}
\begin{split}
\rho(t)=1-\frac{1}{N^2}\sum^N_{i,j=1}\delta_{i,j}(0)exp[-\textbf{P}_j(0)\mathbb{P}\textbf{E}_i^\top],
\end{split}
\end{equation}
where $\delta_{i,j}(0)=0$ for $j=i$, and $\delta_{i,j}(0)=1$ otherwise. Here, $\textbf{P}_j(0)$ represents the supravector of probabilities at time $t=0$ (assuming that the walker starts at node $j$) and the matrix $\mathbb{P}$ indicates the probability of reaching each node through any path of length $1,2,\cdots,t+1$.
Furthermore, $\textbf{E}_i=(\textbf{e}_i,\textbf{e}_i,\cdots,\textbf{e}_i)$ is the supravector in which $\textbf{e}_i$ is the $i$th canonical row vector (see Ref. 35 for details about the derivation of Eq. \ref{method.coverage}).

\clearpage

\begin{table*}\label{table.wiki}
\centering
\begin{center}
\caption{Comparison between the ranking by global PageRank multicentrality in the multilayer network and the ranking by PageRank centrality in the aggregated network.}
\begin{tabular}{p{3.2cm}<{\centering} p{4cm}<{\centering} p{3.8cm}<{\centering} p{3.6cm}<{\centering}}
\hline
Entry & Layer & Global PageRank multicentrality & PageRank centrality\\
\hline
United States & Countries & 1 & 1 (+0) \\
Europe & Geography & 2 & 3 (+1) \\
World War II & History & 3 & 7 (+4) \\
France & Countries & 8 & 2 (-6) \\
Animal & Science & 9 & 77 (+68) \\
Christianity &  Religion & 17 & 19 (+2)\\
Earth & Science & 18 & 39 (+21) \\
20th century & History & 26 & 37 (+11) \\
Agriculture &  Everyday life & 28 & 49 (+21)\\
Television & Design and Technology & 61 & 197 (+136)\\
\hline
\end{tabular}
\end{center}
\end{table*}

\clearpage

\clearpage

\clearpage

\clearpage

\beginsupplement
\clearpage
\section{Supplementary Information}
\subsection{Supplementary Text}
\subsubsection*{I. Tensorial representations}
We first introduce a mathematical formulation of multilayer networks\cite{de2013mathematical}, in which tensorial representations are adopted.
The column vector $\textbf{e}(s)=(0,\ldots,0,1,0,\ldots,0)^\top$ is a canonical vector in the vector space $\mathbb{R}^N$, in which the $s$th component is equal to one and others are zeros.
In a network with $N$ nodes, to describe the relations between any two nodes, the Kronecker product $\otimes$ is adopted. Note that $\textbf{E}(st)=\textbf{e}(s)\otimes\textbf{e}^\top(t)$ is a second-order canonical tensor in the space $\mathbb{R}^{N\times N}$.
Given the intensity $u_{st}$
of the link between nodes $s$ and $t$, the tensor $\textbf{U}\in \mathbb{R}^{N\times N}$ encodes all links  in the network, where
\begin{equation}
\textbf{U}=\sum_{s,t=1}^Nu_{st}\textbf{E}(st).
\end{equation}
Next, we follow the tensorial notation in which $a^i\ (i=1,2,\ldots,N)$ represents the $i$th component of a contravariant vector (column vector) $\textbf{a}\in \mathbb{R}^N$
while $a_j\  (j=1,2,\ldots,N)$ represents the $j$th component of the corresponding covariant vector (row vector) $\textbf{a}^\top\in \mathbb{R}^N$.
We use $(s)$ to indicate the $s$th vector, and $(st)$ to denote the $(st)$-th tensor. Further,  ${i}$ and ${j}$ are used to represent the corresponding components of a certain vector or tensor.
The adjacency tensor $\textbf{U}$ can be written as a 1-covariant and 1-contravariant tensor
\begin{equation}
U_j^i=\sum_{s,t=1}^Nu_{st}e^i(s)e_j(t)=\sum_{s,t=1}^Nu_{st}E^i_j(st),
\end{equation}
where $E^i_j(st)$ indicates the component of $\textbf{E}(st)$ in the $i$th row and $j$th column.

In multilayer networks, there may be various intra- and interlayer links. We further introduce the second-order adjacency tensor $C^i_j(\theta\phi)$ ($\theta,\phi=1,2,\ldots,K$)  to describe them.
To avoid confusion, Latin letters  indicate nodes in networks and Greek letters represent  layers in the multilayer networks, the same as described in the article.
Clearly, $C^i_j(\theta\theta)$ encodes the intralayer links in  layer $\theta$, and $C^i_j(\theta\phi)$ encodes the interlayer links from layer $\theta$ to layer  $\phi$.
Further, $u_{st}(\theta\phi)$ indicates the intensity from node $s$ in  layer $\theta$ to node $t$ in  layer $\phi$.
Because of the varying influence between layers,  we use a canonical basis of layers.
The vector $\textbf{e}^\top(\theta)\in \mathbb{R}^K(\theta=1,2,\ldots,K)$ is the $\theta$th covariant canonical vector  and $e^{\alpha}(\theta)$ represents the $\alpha$th component of $\textbf{e}^\top(\theta)$, while $e_{\beta}(\theta)$ denotes the $\beta$th component of the corresponding contravariant canonical vector $\textbf{e}_\theta$.
The set of all second-order tensors $E^{\alpha}_{\beta}(\theta\phi)=e^{\alpha}(\theta)e_{\beta}(\phi)$ forms the canonical basis of the space $\mathbb{R}^{K\times K}$.
The multilayer adjacency tensor $M^{i\alpha}_{j\beta}$ can be represented by the tensor product of
$C^i_j(\theta\phi)$ and $E^{\alpha}_{\beta}(\theta\phi)$:

\begin{equation*}
\begin{split}
M^{i\alpha}_{j\beta}=&\sum_{\theta,\phi=1}^K C^i_j(\theta\phi)E^{\alpha}_{\beta}(\theta\phi)\\
=&\sum_{\theta,\phi=1}^K\sum_{s,t=1}^Nu_{st}(\theta\phi)\pi^{i\alpha}_{j\beta}(st\theta\phi),
\end{split}
\end{equation*}
where the fourth-order tensor $\pi^{i\alpha}_{j\beta}(st\theta\phi)=e^i(s)e_j(t)e^{\alpha}(\theta)e_{\beta}(\phi)$ forms the canonical basis in the space $\mathbb{R}^{N\times N\times K\times K}$.

Introducing  $w_{{\alpha\beta}}$ to indicate the interlayer influence from  layer $\alpha$ to  layer $\beta$, the tensor $W\in \mathbb{R}^{K\times K}$ encodes the interlayer influence between any two layers in  multilayer networks, where $W$ can be written as the 1-covariant and 1-contravariant tensor
\begin{equation}
W^{\alpha}_{\beta}=\sum_{\theta,\phi=1}^K w_{{\alpha\beta}}e^{\alpha}(\theta)e_{\beta}(\phi)=\sum_{\theta,\phi=1}^K w_{{\alpha\beta}}E^{\alpha}_{\beta}(\theta\phi).
\end{equation}
Finally, we introduce the fourth-order tensor $H$, called the interaction tensor,  in terms of the  tensor product
\begin{equation*}
\begin{split}
H^{i\alpha}_{j\beta}=&W^{\alpha}_{\beta}\cdot M^{i\alpha}_{j\beta}\\
=&\sum_{\theta,\phi=1}^K\sum_{s,t=1}^Nu_{st}(\theta\phi)w_{\alpha\beta}\pi^{i\alpha}_{j\beta}(st\theta\phi),
\end{split}
\end{equation*}
which encodes the interactions between any two nodes in the multilayer networks.

To simplify the notation, the Einstein summation convention is  adopted. Repeated indices in a term, where one index is a subscript and the other is a superscript, denote the summation  over all indices, which is also called  a contraction in tensorial operations. For example, we use the convention in the following equations:
\begin{equation*}
\begin{split}
A_i^i&=\Sigma_{i=1}^NA_i^i,\\
A_j^i B^j_i&=\Sigma_{i=1}^N\Sigma_{j=1}^NA_j^i B^j_i,\\
A_{j\beta}^{i\alpha} B_{i\alpha}&=\Sigma_{i=1}^N\Sigma_{\alpha=1}^K A_{j\beta}^{i\alpha} B_{i\alpha}.
\end{split}
\end{equation*}
This convention can be adopted for the product of any number of tensors with arbitrary order.

\subsubsection*{II. Tensor decomposition}\label{ap.tensordecomposition}
Tensors can be decomposed into lower-dimensional objects like matrices and supravectors\cite{kolda2009tensor}. We first flatten the adjacency tensor $M$ into a matrix, i.e., we represent the fourth-order tensor $M\in\mathbb{R}^{N\times N\times K\times K}$ by a matrix $\overline{M}\in \mathbb{R}^{NK\times NK}$.
The overline here allows us to distinguish a tensor from its lower-dimensional form, such as a matrix.
Thus, the matrix $\overline{M}$ is a block matrix
\begin{equation}
\overline{M}= \begin{pmatrix}
\begin{BMAT}(@,20pt,20pt){c.c.c.c}{c.c.c.c}
\overline{M}_{11} & \overline{M}_{12} & \cdots & \overline{M}_{1K}  \\
\overline{M}_{21} & \overline{M}_{22} & \cdots & \overline{M}_{2K}  \\
\vdots & \vdots & \ddots & \vdots  \\
\overline{M}_{K1} & \overline{M}_{K2} & \cdots & \overline{M}_{KK}  \\
\end{BMAT}
\end{pmatrix},
\end{equation}
where the block $\overline{M}_{\alpha\beta}\in \mathbb{R}^{N\times N}$ encodes the links from  layer $\alpha$ to layer $\beta$.

We can vectorize the second-order tensor $\Phi_{i\alpha}\in\mathbb{R}^{N\times K}$ into a supravector $\overline{\Phi}\in \mathbb{R}^{NK}$:
\begin{equation}
\overline{\Phi}= \begin{pmatrix}
\begin{BMAT}(@,20pt,20pt){c}{c.c.c.c}
\overline{\Phi}_{1}  \\
\overline{\Phi}_{2}  \\
\vdots  \\
\overline{\Phi}_{K}  \\
\end{BMAT}
\end{pmatrix},
\end{equation}
where $\overline{\Phi}_\alpha\in \mathbb{R}^N$ encodes the multicentrality of nodes in  layer $\alpha$.

Denoting the matrix form of the influence tensor $W$ by $\overline{W}\in \mathbb{R}^{K\times K}$, we have
\begin{equation}
\overline{W}= \begin{pmatrix}
\begin{BMAT}(@,20pt,20pt){c.c.c.c}{c.c.c.c}
w_{11} & w_{12} & \cdots & w_{1K}  \\
w_{21} & w_{22} & \cdots & w_{2K}  \\
\vdots & \vdots & \ddots & \vdots  \\
w_{K1} & w_{K2} & \cdots & w_{KK}  \\
\end{BMAT}
\end{pmatrix},
\end{equation}
where element $w_{\alpha\beta}$ quantifies the interlayer influence from layer $\alpha$ to  layer $\beta$.
Using these decompositions, the tensorial equation
\begin{equation}\label{eq.simain}
\begin{split}
H_{j\beta}^{i\alpha}\Phi_{i\alpha}=\lambda_\beta\Phi_{j\beta}
\end{split}
\end{equation}
becomes a  matrix equation

\begin{equation}\label{eq.mainmatrix}
\begin{split}
&\begin{pmatrix}
\begin{BMAT}(@,20pt,20pt){c.c.c.c}{c.c.c.c}
w_{11}\cdot\overline{M}_{11} & w_{12}\cdot\overline{M}_{12} & \cdots & w_{1K}\cdot\overline{M}_{1K}  \\
w_{21}\cdot\overline{M}_{21} & w_{22}\cdot\overline{M}_{22} & \cdots & w_{2K}\cdot\overline{M}_{2K}  \\
\vdots & \vdots & \ddots & \vdots  \\
w_{K1}\cdot\overline{M}_{K1} & w_{K2}\cdot\overline{M}_{K2} & \cdots & w_{KK}\cdot\overline{M}_{KK}  \\
\end{BMAT}
\end{pmatrix}^\top\cdot\begin{pmatrix}
\begin{BMAT}(@,20pt,20pt){c}{c.c.c.c}
\overline{\Phi}_{1}  \\
\overline{\Phi}_{2}  \\
\vdots  \\
\overline{\Phi}_{K}  \\
\end{BMAT}
\end{pmatrix}= \begin{pmatrix}
\begin{BMAT}(@,20pt,20pt){c}{c.c.c.c}
\lambda_{1}\cdot\overline{\Phi}_{1}  \\
\lambda_{2}\cdot\overline{\Phi}_{2}  \\
\vdots  \\
\lambda_{K}\cdot\overline{\Phi}_{K}  \\
\end{BMAT}
\end{pmatrix}.
\end{split}
\end{equation}

We further define the vector $\Lambda=[\lambda_{1},\lambda_{2},\cdots,\lambda_{K}]^\top \in\mathbb{R}^{K}$. Then the Eq. \ref{eq.mainmatrix} can be written as
\begin{equation}
(\overline{W}\odot\overline{M})^\top\cdot\overline{\Phi}=\Lambda\odot\overline{\Phi},
\end{equation}
where $\odot$ denotes the Khatri-Rao product.

In the calculation of the global multicentrality in multilayer networks, we take
\begin{equation}
w_{\alpha\beta}=f(\overline{\Phi}_{\alpha})/f(\overline{\Phi}_{\beta}),
\end{equation}
and we prove that there exists a unique normalized supravector $\overline{\Phi}$ satisfying
\begin{equation}
(\overline{W}\odot\overline{M})^\top\cdot\overline{\Phi}=\lambda_1\cdot\overline{\Phi},
\end{equation}
where $\lambda_{\alpha}=\lambda_1$\ ($\alpha=1,2,\cdots,K$).

In the calculation of the local multicentrality in multilayer networks, we take
\begin{equation}
\begin{split}
w_{\alpha\beta}=\frac{\sum_{i,j=1}^NM^{j\beta}_{i\alpha}\Phi_{j\beta}}{\sum_{i,j=1}^NM^{i\alpha}_{j\beta}\Phi_{i\alpha}},
\end{split}
\end{equation}
and we also prove that there exists  a unique normalized supravector $\overline{\Phi}$ satisfying
\begin{equation}
(\overline{W}\odot\overline{M})^\top\cdot\overline{\Phi}=\Lambda\odot\overline{\Phi}.
\end{equation}

It is possible that different layers have different numbers of nodes, resulting in different tensor size. To maintain the unified tensor framework, we need to expand the tensor $M$ and $\Phi$ to a specific size by adding zeros for nodes which are not in the current layer.
Thus, after tensor decomposition, many rows and the corresponding columns only have zero elements in the matrix $\overline{M}$, and the corresponding multicentrality $\Phi_{i\alpha}$ are also zeros. To reduce the computational complexity in practice, we can omit these rows and columns in the matrix $\overline{M}$ and the corresponding elements in the supravector $\overline{\Phi}$.

\subsubsection*{III. PageRank centrality in a single-layer network}\label{ap.pagerank}
Consider a graph $G=(V,E)$, where $V=\{v_1,v_2,\cdots,v_N\}$ is the set of nodes and $E$ is the set of  directed links. We have the adjacency matrix $A=(a_{ij})$, where
\begin{equation}
a_{ij}=
\begin{cases}
1,& \text{if $(i,j)\in E$}\\
0,& \text{otherwise}
\end{cases}.
\end{equation}
Further, we construct a stochastic matrix $T=(t_{ij})$, which is the transposition of the adjacency matrix $A$ after normalization, i.e.,
\begin{equation}
t_{ij}=
\begin{cases}
\frac{1}{k^{out}_j},& \text{if $(j,i)\in E$}\\
0,& \text{otherwise}
\end{cases}
\end{equation}
where $k^{out}_j$ is the out-degree  of node $j$. The PageRank  centrality $C\in \mathbb{R}^N$\cite{brin2012reprint} is the normalized leading right eigenvector of $T$, which corresponds to the leading eigenvalue $\lambda_1=1$, i.e., 
\begin{equation}
\begin{split}
T\cdot C=1\cdot C.
\end{split}
\end{equation}

Next, we will briefly prove the existence and uniqueness of $C$.
We first assume that the graph $G$ is strongly connected. 
Clearly, the summation of each column of $T$ is equal to one, and $T$ has a left eigenvector $\textbf{1}_{1\times N}$, whose components are all equal to one, corresponding to the case of eigenvalue $\lambda=1$.  Since $\sum_i|t_{ij}|=1$, all eigenvalues satisfy $|\lambda_j|\leq1$ according to Gershgorin's circle theorem\cite{berman1994nonnegative}.
Furthermore,  the matrix $T$ is irreducible since the graph $G$ is strongly connected. Hence, the leading eigenvalue $\lambda_1$ of $T$ is unique  according to the Perron-Frobinus theorem\cite{gantmakher1998theory}, leading to the fact that the normalized leading eigenvector $C$ is unique.

When the graph $G$ is not strongly connected, the following two steps can be employed to guarantee  the existence and uniqueness of the PageRank centrality. At the first step, find all dead ends (nodes with no outgoing links) in the graph $G$, and add outgoing links from them to all nodes.   Since we add outgoing links between a dead end and all nodes in the graph, such an operation will not change the relative centrality of other nodes\cite{agarwal2006learning}. At the second step, introduce a damping factor $d$ ranging between zero and one\cite{boldi2005pagerank}, and  the corresponding stochastic matrix $T_{d}$ becomes
\begin{equation}
T_{d}=d\cdot T+\frac{1-d}{N}\cdot\textbf{1}_{N\times N}.
\end{equation}
This makes  $T_d$ irreducible, and thus the leading eigenvalue $\lambda_1$ of $T$ is unique\cite{gantmakher1998theory}.  

The PageRank centrality can be interpreted by two models: the flow model\cite{agarwal2006learning} and the random walk model\cite{chen2007finding}.
In the flow model,  the PageRank score is regarded as a kind of negotiable flow.
At each iteration, the current flow (score) of each node is distributed through its outgoing links on an equal footing to its neighboring nodes.
In the steady state, the flow that each node gains is its PageRank score. In the random walk model, we consider that a random walker  starts at a certain node and randomly chooses an outgoing link to walk iteratively.
In this context, the PageRank score of each node is the probability of the  walker resting on the node in the steady state.

\subsubsection*{IV. PageRank multicentrality in multilayer networks}\label{ap.pagerankmul}
In multilayer networks, we define the stochastic tensor $T_{j\beta}^{i\alpha}=\frac{M_{j\beta}^{i\alpha}}{\sum_{j,\beta}M_{j\beta}^{i\alpha}}$ such that $v^{j\beta}T_{j\beta}^{i\alpha}=v^{i\alpha}$, where $v$ is the tensor with all components equal to one.
Further, we define the interaction tensor as $H_{j\beta}^{i\alpha}=W^\alpha_{\beta}\cdot T_{j\beta}^{i\alpha}$. It follows from Eq. \ref{eq.simain} that the leading eigentensor of $H$, denoted by $\Phi_{i\alpha}$, is  the PageRank multicentrality of node $i$ in layer $\alpha$, i.e.,
\begin{equation}
H_{j\beta}^{i\alpha}\Phi_{i\alpha}=1\cdot\Phi_{j\beta}.
\end{equation}
Also, the influence tensor $W$ can be a function of the tensor $T$ and centrality $\Phi$.
It is noteworthy that in PageRank multicentrality, the leading eigenvalue $\lambda_1$ of the interaction tensor is equal to one in accordance with PageRank centrality in a single-layer network.
As before, a damping factor $d \in [0,1]$ should be introduced to guarantee the existence and uniqueness of global and local PageRank multicentrality when the multilayer networks are not strongly connected (see Text III of the SI above for further details).

Similarly, to characterize global PageRank multicentrality, we can choose the influence tensor as $W^\alpha_{\beta}=f(\Phi_{:\alpha})/f(\Phi_{:\beta})$.
In particular, after a node in layer $\alpha$ equally distributes its multicentrality score to its neighbors, the scores propagating along intralayer links stay unchanged, while the scores propagating along interlayer links to nodes in layer $\beta$ will be scaled by a factor $f(\Phi_{:\alpha})/f(\Phi_{:\beta})$.
In this case, we prove that the leading eigenvalue of $H$ is equal to $1$ (see Text V of the SI below for further details about the theoretic analysis).
For clarification, we provide  a toy example in Supplementary Fig. \ref{fig.pagepaper}.

In local PageRank multicentrality, we specify the influence tensor as \begin{equation}\label{eq.prlocal}
\begin{split}
W^\alpha_{\beta}=\frac{\sum_{i,j=1}^NT^{j\beta}_{i\alpha}\Phi_{j\beta}}{\sum_{i,j=1}^NT^{i\alpha}_{j\beta}\Phi_{i\alpha}}.
\end{split}
\end{equation}
Likewise, we prove that the leading eigenvalue of $H$ remains one (see Text VI of the SI below for the proof).
When a node in layer $\alpha$ propagates its multicentrality score to nodes in layer $\beta$, the interactions depend on the scores from nodes in layer $\beta$  to nodes in  layer $\alpha$.
In other words, the scores leaving from layer $\alpha$ will be returned to layer $\alpha$. We provide an illustration of local PageRank multicentrality in Supplementary Fig. \ref{fig.local}.
In the random walk model, the interaction  tensor $H^{i\alpha}_{j\beta}$ encodes the transition probability of node $i$ in layer $\alpha$ leaving for any node $j$ in layer $\beta$.
We can interpret the local multicentrality by a random walk process over a multilayer network as follows.
If a walker starts at a certain node in layer $\alpha$, she will walk along the intralayer links normally according to the transition probability.
However, once the walker  leaves from  layer $\alpha$ to any other layer along the interlayer links, she will not rest on nodes in other layers until she goes back to layer $\alpha$. The local PageRank multicentrality score of each node in layer $\alpha$ is the probability of such a walker resting on the node in the steady state.

\subsubsection*{V. Theoretical demonstration of global multicentrality}\label{ap.globalproof}

\textit{Theorem 1:} Given a strongly connected  multilayer network $\mathcal{M}$ and a function $f:\mathbb{R}^N_+\rightarrow\mathbb{R}_+$, we denote the eigenvector centrality in the corresponding single-layer network by $\Theta$. Then, a unique global eigenvector multicentrality $\Phi$ exists if and only if, for $\forall\ \Theta_{:\alpha}, \Theta_{:\beta}\ (\alpha,\beta=1,2,\cdots,K)$, there exists only one $y_{\alpha\beta}\in (0,+\infty)$ such that $f(y_{\alpha\beta}\cdot \Theta_{:\beta})=1/y_{\alpha\beta}\cdot f(\Theta_{:\alpha})$.

\textit{Proof of Theorem 1.}
(Sufficiency.) First, we will prove the existence of $\Phi$.
Because the multilayer network is strongly connected,  there is a unique normalized leading eigentensor $\Theta$ (i.e., the eigenvector centrality in the corresponding single-layer network), satisfying
\begin{equation}
\begin{split}
&M_{j\beta}^{i\alpha}\Theta_{i\alpha}=\lambda_{1}\Theta_{j\beta},
\end{split}
\end{equation}
which we have stated in Text III.
Then for $\Theta_{:1}\in \mathbb{R}^N$ and $\Theta_{:\gamma}\in \mathbb{R}^N (\gamma =1,2,\cdots,K)$, there exists only one $y_{1\gamma} \in (0,+\infty)$ such that $f(y_{1\gamma}\cdot \Theta_{:\gamma})=1/y_{1\gamma}\cdot f(\Theta_{:1})$.
Let $\Phi_{:\gamma}=y_{1\gamma}\cdot \Theta_{:\gamma}\ (\gamma=1,2,\cdots,K)$. We have
\begin{equation*}
\begin{split}
\sum_{\alpha=1}^{K}\sum_{i=1}^{N}H^{i\alpha}_{j\beta} \Phi_{i\alpha}=&\sum_{\alpha=1}^{K}\sum_{i=1}^{N}(w_{\alpha\beta} M^{i\alpha}_{j\beta}) \cdot (y_{1\alpha} \cdot \Theta_{i\alpha})\\
=&y_{1\beta}\sum_{\alpha=1}^{K}\sum_{i=1}^{N} M^{i\alpha}_{j\beta}\Theta_{i\alpha}\\
=&\lambda_1\Phi_{j\beta}.
\end{split}
\end{equation*}

Then, we will prove the uniqueness of $\Phi$ by contradiction. Suppose there exist two tensors $\Phi'$ and $\Phi''$ such that $\sum_{i,\alpha}\Phi'_{i\alpha}=\sum_{i,\alpha}\Phi''_{i\alpha}$ and
\begin{equation*}
\begin{split}
\begin{cases}
H_{j\beta}^{i\alpha}\Phi'_{i\alpha}=\lambda_1\Phi'_{j\beta},\\
H_{j\beta}^{i\alpha}\Phi''_{i\alpha}=\lambda_1\Phi''_{j\beta}.
\end{cases}
\end{split}
\end{equation*}
For $\Phi'_{i\alpha}$, we have
\begin{equation*}
\begin{split}
\sum_{i,\alpha}H_{j\beta}^{i\alpha}\Phi'_{i\alpha}&=\lambda_{1}\cdot \Phi'_{j\beta}\\
\sum_{i,\alpha}\frac{f(\Phi'_{:\alpha})}{f(\Phi'_{:\beta})}M_{j\beta}^{i\alpha}\Phi'_{i\alpha}&=\lambda_{1}\cdot \Phi'_{j\beta}\\
\sum_{i,\alpha}M_{j\beta}^{i\alpha}\cdot f(\Phi'_{:\alpha})\Phi'_{i\alpha}&=\lambda_{1}\cdot f(\Phi'_{:\beta})\Phi'_{j\beta}.
\end{split}
\end{equation*}
Similarly for $\Phi''_{:\alpha}$, we  have
\begin{equation*}
\begin{split}
\sum_{i,\alpha}M_{j\beta}^{i\alpha}\cdot f(\Phi''_{:\alpha})\Phi''_{i\alpha}&=\lambda_{1}\cdot f(\Phi''_{:\beta})\Phi''_{j\beta}.
\end{split}
\end{equation*}
Due to the uniqueness of the leading eigentensor $\Theta$ of $M$, we have $f(\Phi'_{:\alpha})\cdot \Phi'_{i\alpha}=f(\Phi''_{:\alpha})\cdot\Phi''_{i\alpha}=k\cdot\Theta_{:\alpha}$, and  $\Phi'=\Phi''$. This is in contradiction with our hypothesis. Thus, the solution $\Phi$ is unique.

(Necessity.) We proceed to prove the necessity of Theorem by contradiction.
Consider a strongly connected  multilayer network $\mathcal{M}$ and a function $f:\mathbb{R}^N_+\rightarrow\mathbb{R}_+$. Suppose that a unique global eigenvector multicentrality $\Phi$ exists. Further, we suppose that
$\exists\ \Theta_{:\mu},\Theta_{:\nu}$, and either of the following two conditions is satisfied: 1) $\nexists\ y_{\mu\nu}$ such that $f(y_{\mu\nu}\cdot \Theta_{:\nu})=1/y_{\mu\nu}\cdot f(\Theta_{:\mu})$; 2)  $\exists$ $y'_{\mu\nu},y''_{\mu\nu}\in (0,+\infty)$ such that $f(y'_{\mu\nu}\cdot \Theta_{:\nu})=1/y'_{\mu\nu}\cdot f(\Theta_{\mu})$ and $f(y''_{\mu\nu}\cdot \Theta_{\nu})=1/y''_{\mu\nu}\cdot f(\Theta_{\mu})$.

By hypothesis, there is  a unique global eigenvector multicentrality $\Phi\in\mathbb{R}^{N\times K}$ for the tensor $M$, such that
\begin{equation}
\begin{split}
H_{j\beta}^{i\alpha}\Phi_{i\alpha}=\lambda_1\Phi_{j\beta},
\end{split}
\end{equation}
i.e.,
\begin{equation*}
\begin{split}
\sum_{i,\alpha}H_{j\beta}^{i\alpha}\Phi_{i\alpha}&=\lambda_{1}\cdot \Phi_{j\beta}\\
\sum_{i,\alpha}\frac{f(\Phi_{:\alpha})}{f(\Phi_{:\beta})}M_{j\beta}^{i\alpha}\Phi_{i\alpha}&=\lambda_{1}\cdot \Phi_{j\beta}\\
\sum_{i,\alpha}M_{j\beta}^{i\alpha}\cdot f(\Phi_{:\alpha})\Phi_{i\alpha}&=\lambda_{1}\cdot f(\Phi_{:\beta})\Phi_{j\beta}.
\end{split}
\end{equation*}
Then we have

\begin{equation*}
\begin{cases}
f(\Phi_{:\mu})\cdot\Phi_{:\mu}=k\cdot\Theta_{:\mu}\\
f(\Phi_{:\nu})\cdot\Phi_{:\nu}=k\cdot\Theta_{:\nu}.
\end{cases}
\end{equation*}
Further, there exist $y_{\mu\mu}>0$ and $y_{\mu\nu}>0$ such that $\frac{y_{\mu\mu}}{y_{\mu\nu}}$ is a constant, and the following two equations hold: 
\begin{equation*}
\begin{cases}
\Phi_{:\mu}=y_{\mu\mu}\cdot\Theta_{:\mu}\\
\Phi_{:\nu}=y_{\mu\nu}\cdot\Theta_{:\nu}.
\end{cases}
\end{equation*}
Without loss of generality, we let $y_{\mu\mu}=1$, i.e., $\Phi_{:\mu}=\Theta_{:\mu}$.
Thus, there exists only one $y_{\mu\nu}>0$ such that
\begin{equation}
f(\Phi_{:\mu})=f(\Phi_{:\nu})\cdot y_{\mu\nu}=k,
\end{equation}
i.e.,
\begin{equation}
f(y_{\mu\nu}\cdot\Theta_{:\nu})=1/y_{\mu\nu}\cdot f(\Theta_{:\mu}),
\end{equation}
which is in contradiction with the two conditions of the hypothesis.
$\blacksquare$

Note that $\Phi_{:\alpha}$ represents the eigenvector multicentrality scores of nodes in layer $\alpha$, $\Theta_{:\beta}$ represents the eigenvector centrality scores of nodes in layer $\beta$, and $\Phi_{:\beta}=y_{\alpha\beta}\cdot\Theta_{:\beta}$ represents the eigenvector multicentrality of nodes in layer $\beta$. Thus, $y_{\alpha\beta}$ is a scale factor comparing the eigenvector multicentrality in a multilayer network with the eigenvector centrality in the corresponding single-layer network. If $y_{\alpha\beta}<1$, the eigenvector multicentrality score of nodes in layer $\beta$ is relatively less important than the eigenvector centrality score in the corresponding single-layer network when taking layer $\alpha$ as a reference ($\Phi_{:\alpha}=\Theta_{:\alpha}$), and vice versa.
Hence, it is necessary to guarantee the existence and  uniqueness of $y_{\alpha\beta}$ for any $\Theta_{:\alpha}$ and $\Theta_{:\beta}$ as the theorem states. In fact, the condition in the theorem requires that the slope of the function $f$  has a lower bound. So, if the function $f$ is a nonnegative monotonic increasing function such as the $L^p$-norm, the condition in the theorem is satisfied since the slope of  a  monotonic increasing function is always positive.

In addition, from the proof above, we can see that $\lambda_1$ is the leading eigenvalue of the tensor $M$, satisfying
\begin{equation}
\begin{split}
&M_{j\beta}^{i\alpha}\Theta_{i\alpha}=\lambda_{1}\Theta_{j\beta}.
\end{split}
\end{equation}
Also, $\lambda_1$ is the leading eigenvalue of the tensor $H$, satisfying
\begin{equation}
\begin{split}
H_{j\beta}^{i\alpha}\Phi_{i\alpha}=\lambda_1\Phi_{j\beta}.
\end{split}
\end{equation}

In the global PageRank multicentrality, the leading eigenvalue of the tensor $M$ is equal to one. Clearly, the leading eigenvalue of the tensor $H$ is also equal to one.

\subsubsection*{VI. Theoretical demonstration of local multicentrality}\label{ap.localproof}

\textit{Theorem 2:} Given a strongly connected multilayer network $\mathcal{M}$ and influence matrix $\overline{W}$ described by
\begin{equation}
\begin{split}
w_{\alpha\beta}=\frac{\sum_{i,j=1}^NM^{j\beta}_{i\alpha}\Phi_{j\beta}}{\sum_{i,j=1}^NM^{i\alpha}_{j\beta}\Phi_{i\alpha}},
\end{split}
\end{equation}
the  local eigenvector multicentrality $\Phi$ has a unique normalized solution to the equation
\begin{equation}\label{eq.siframe}
\begin{split}
&H_{j\beta}^{i\alpha}\Phi_{i\alpha}=\lambda_{\beta}\Phi_{j\beta}.
\end{split}
\end{equation}

\textit{Proof of Theorem 2.}
We  flatten the interaction tensor $H$ into a matrix $\overline{H}\in \mathbb{R}^{NK\times NK}$, and vectorize the second-order tensor $\Phi\in\mathbb{R}^{N\times K}$ into a vector $\overline{\Phi}\in \mathbb{R}^{NK}$.
We will show that  $\overline{H}\cdot\overline{\Phi}$ is a contraction mapping of $\overline{\Phi}$. To this end, we first suppose that the adjacency tensor $M$ satisfies $v^{j\beta}M_{j\beta}^{i\alpha}=v^{i\alpha}$, where $v$ is the tensor with all components equal to one, i.e., the case of local PageRank multicentrality.
Consider two different vectors $\overline{\Phi}'_\gamma$ and $\overline{\Phi}''_\gamma$. For any given $\overline{\Phi}_\eta$ ($\eta\neq\gamma$), we suppose
\begin{equation}
\overline{\Phi}'(\gamma)= \begin{pmatrix}
\begin{BMAT}(@,20pt,20pt){c}{c.c.c.c.c.c}
\overline{\Phi}_{1}  \\
\overline{\Phi}_{2}  \\
\vdots  \\
\overline{\Phi}'_{\gamma}\\
\vdots  \\
\overline{\Phi}_{M}  \\
\end{BMAT}
\end{pmatrix}\in \mathbb{R}^N,  \
\overline{\Phi}''(\gamma)= \begin{pmatrix}
\begin{BMAT}(@,20pt,20pt){c}{c.c.c.c.c.c}
\overline{\Phi}_{1}  \\
\overline{\Phi}_{2}  \\
\vdots  \\
\overline{\Phi}''_{\gamma}\\
\vdots  \\
\overline{\Phi}_{M}  \\
\end{BMAT}
\end{pmatrix}\in \mathbb{R}^N.
\end{equation}
Then

\begin{align*}
&|\overline{H}\cdot\overline{\Phi}'(\gamma)-\overline{H}\cdot\overline{\Phi}''(\gamma)|\\
=&|\overline{M}_{\gamma\gamma}\overline{\Phi}'_{\gamma}+\sum_{\eta\neq\gamma}\frac{\|\overline{M}_{\eta\gamma}\cdot \overline{\Phi}'_{\eta}\|_1}{\|\overline{M}_{\gamma\eta}\cdot \overline{\Phi}'_{\gamma}\|_1}\cdot \overline{M}_{\gamma\eta}\cdot \overline{\Phi}'_{\gamma}-\overline{M}_{\gamma\gamma}\overline{\Phi}''_{\gamma}-\sum_{\eta\neq\gamma}\frac{\|\overline{M}_{\eta\gamma}\cdot \overline{\Phi}''_{\eta}\|_1}{\|\overline{M}_{\gamma\eta}\cdot \overline{\Phi}''_{\gamma}\|_1}\cdot \overline{M}_{\gamma\eta}\cdot \overline{\Phi}''_{\gamma}|\\
\leqslant &p_\gamma\cdot |\overline{\Phi}'_{\gamma}-\overline{\Phi}''_{\gamma}|,
\end{align*}
where $p_\gamma$ is given by
\begin{equation}
p_\gamma=\max_{i,k}\sum_j\|\overline{M}_{\gamma\gamma}(i,j)-\overline{M}_{\gamma\gamma}(k,j)\|_1<1.
\end{equation}
Thus, for any positive vectors $\overline{\Phi}'$ and $\overline{\Phi}''$,
\begin{equation*}
\begin{split}
|\overline{H}\cdot \overline{\Phi}'-\overline{H}\cdot \overline{\Phi}''|\leqslant  \max_\gamma \{p_\gamma\}\cdot |\overline{\Phi}'-\overline{\Phi}''|.
\end{split}
\end{equation*}
This implies that the function $\overline{H}\cdot \overline{\Phi}$ is a contraction mapping of $\overline{\Phi}$. According to the Banach fixed-point theorem\cite{granas2013fixed}, $\overline{H}$ admits a unique fixed-point $\overline{\Phi}$, satisfying
\begin{equation}
\begin{split}
&\overline{H}\cdot\overline{\Phi}=\overline{\Phi},
\end{split}
\end{equation}
from which we find that the eigenvalue of $H_{j\beta}^{i\alpha}$ remains one in the local PageRank multicentrality analysis.
For the general case, we can construct an analogous contraction mapping.
$\blacksquare$

\subsubsection*{VII. Epidemic spreading in  multilayer networks}\label{Epidemic spreading}

We first introduce epidemic spreading process in a single-layer network. There are many models describing epidemic  spreading process, such as the susceptible-infected (SI) model, the susceptible-infected-susceptible (SIS) model, the susceptible-infected-recovered (SIR) model and so on\cite{bailey1975mathematical}. Here we take the SI model as an illustration.

In an undirected single-layer  social network, nodes denote individuals and links indicate their social relations. We study the spreading of a disease in such a social network.
Each node of the network has only  two states, ``susceptible'' (S) or ``infected'' (I). A susceptible node is a temporarily healthy node, which can be infected by infected nodes, while an infected node will stay in the infected state.
At each step, an infected node will infect the susceptible nodes connected to it with probability $\mu$, which is called the transition rate. Clearly, if a susceptible node has more infected neighbors, it will have a higher probability of being infected.
Since the infected nodes will not  recover, all nodes will eventually  be infected  if the network is strongly connected and $\mu>0$.

In multilayer networks, nodes lie in different layers. The transition rate can be different along intralayer links and interlayer links. Here we consider a dataset  from a large European research institute\cite{yin2017local} (data provided by the Stanford Network Analysis Project, \url{http://snap.stanford.edu/index.html}). There are 1,005 individuals, each belonging to exactly one of 42 departments at the research institute. A multilayer network is built with  $K=42$ layers indicating departments, and $N=1,005$ nodes representing individuals. There is an undirected link between two nodes if either one of them has sent at least one email to the other, and the connections are encoded by the adjacency tensor $M \in \mathbb{R}^{N\times N\times K\times K}$.
We then consider an epidemic spreading process in this multilayer network.
Considering a certain node as the infection source, we set the intralayer transition rate to 0.05 and the interlayer transition rate to 0.03 at each  step\cite{pastor2015epidemic}.  Since we already have prior knowledge about interlayer influence, the influence tensor $W \in \mathbb{R}^{K\times K}$ is given as follows:
\begin{equation}
W^{\alpha}_{\beta}=
\begin{cases}
0.05,& \text{$\alpha=\beta$}\\
0.03,& \text{$\alpha\neq\beta$}
\end{cases}.
\end{equation}

Then, we obtain the interaction tensor $H \in \mathbb{R}^{N\times N\times K\times K}$, and the multicentrality tensor $\Phi\in \mathbb{R}^{N\times K}$ by solving the tensorial equation
\begin{equation}
\begin{split}
H_{j\beta}^{i\alpha}\Phi_{i\alpha}=\lambda_1\Phi_{j\beta}.
\end{split}
\end{equation}

We simulate the spreading process in this multilayer network with only one infected node in the initial state. At each  step, the infected node/nodes  will infect its/their susceptible neighboring nodes at the transition rate, and the rates are different between intralayer links and interlayer links as we mentioned above.
All nodes will be infected as long as there is a path between any two nodes in the multilayer network. We focus on the spreading coverage, i.e., the percentage of the infected nodes after each step.
The results are shown in Supplementary Fig. \ref{fig.spreading}, indicating that the nodes with higher eigenvector multicentrality play a more important role in the process of epidemic spreading in a multilayer network.

\subsubsection*{VIII. Theoretical demonstration of the numerical solution}\label{ap.numericalproof}

Here we prove that the iteration scheme in the Materials and Methods section,
\begin{equation}\label{ap.scheme}
\overline{\Phi}^{(k+1)}=\overline{\Phi}^{(k)}+D^{(k)}[(\overline{W}\odot\overline{M})^{\top(k)}\odot{\Omega^{(k)}}-E_N]\overline{\Phi}^{(k)},
\end{equation}
converges to the solution to the matrix equation
\begin{equation}\label{si.matrixequation}
(\overline{W}\odot\overline{M})^\top\cdot\overline{\Phi}=\Lambda\odot\overline{\Phi}.
\end{equation}

\textit{Proof.} We first consider the global multicentrality. The iteration scheme Eq. \ref{ap.scheme} becomes
\begin{equation}\label{ap.globalscheme}
\overline{\Phi}^{(k+1)}=\overline{\Phi}^{(k)}+D^{(k)}[(\overline{W}\odot\overline{M})^{\top(k)}/{\lambda_1^{(k)}}-E_N]\overline{\Phi}^{(k)},
\end{equation}
and the matrix Eq. \ref{si.matrixequation} becomes
\begin{equation}\label{globalequation}
(\overline{W}\odot\overline{M})^\top\cdot\overline{\Phi}=\lambda_1\odot\overline{\Phi}.
\end{equation}

In Text V of the SI above, we have proved that there exists a unique solution to Eq. \ref{globalequation}.
We denote the unique solution by $\overline{\Phi}^*$ and the corresponding eigenvalue by $\lambda_1^*$.
The convergence of the iteration scheme Eq. \ref{ap.globalscheme} has been demonstrated in the Materials and Methods section. We denote the limit of $\overline{\Phi}^{(k)}$ by $\overline{\Phi}^{(\infty)}$. The primary task is to prove that  $\overline{\Phi}^{(\infty)}$ is equal to $\overline{\Phi}^*$.
Clearly,
\begin{equation*}
\begin{split}
\overline{\Phi}^{(\infty)}&=\overline{\Phi}^{(\infty)}+D^{(\infty)}[(\overline{W}\odot\overline{M})^{\top(\infty)}/{\lambda_1^{(\infty)}}-E_N]\overline{\Phi}^{(\infty)}\\
0&=D^{(\infty)}[(\overline{W}\odot\overline{M})^{\top(\infty)}/{\lambda_1^{(\infty)}}-E_N]\overline{\Phi}^{(\infty)}.
\end{split}
\end{equation*}
As the matrix $D^{(\infty)}$ is a diagonal matrix with non-zero diagonal elements, it is invertible. Thus
\begin{equation*}
\begin{split}
[(\overline{W}\odot\overline{M})^{\top(\infty)}/{\lambda_1^{(\infty)}}-E_N]\overline{\Phi}^{(\infty)}&=0\\
(\overline{W}\odot\overline{M})^{\top(\infty)}/{\lambda_1^{(\infty)}}\cdot\overline{\Phi}^{(\infty)}&=\overline{\Phi}^{(\infty)}.
\end{split}
\end{equation*}
Therefore, we have
\begin{equation*}
\begin{split}
(\overline{W}\odot\overline{M})^{\top(\infty)}\cdot\overline{\Phi}^{(\infty)}={\lambda_1^{(\infty)}}\cdot\overline{\Phi}^{(\infty)}.
\end{split}
\end{equation*}
Note that $\overline{\Phi}^{(\infty)}$ satisfies Eq. \ref{globalequation} and has a unique solution $\overline{\Phi}^{*}$. Thus,  we have $\overline{\Phi}^{(\infty)}=\overline{\Phi}^{*}$ and $\lambda_1^{(\infty)}=\lambda_1^*$. Similarly, we can prove the convergence of the iteration scheme for local multicentrality. $\blacksquare$

\clearpage
\subsection{Supplementary Figures}
$\ $

\begin{figure*}[!h]
\centering
\includegraphics[width=\textwidth]{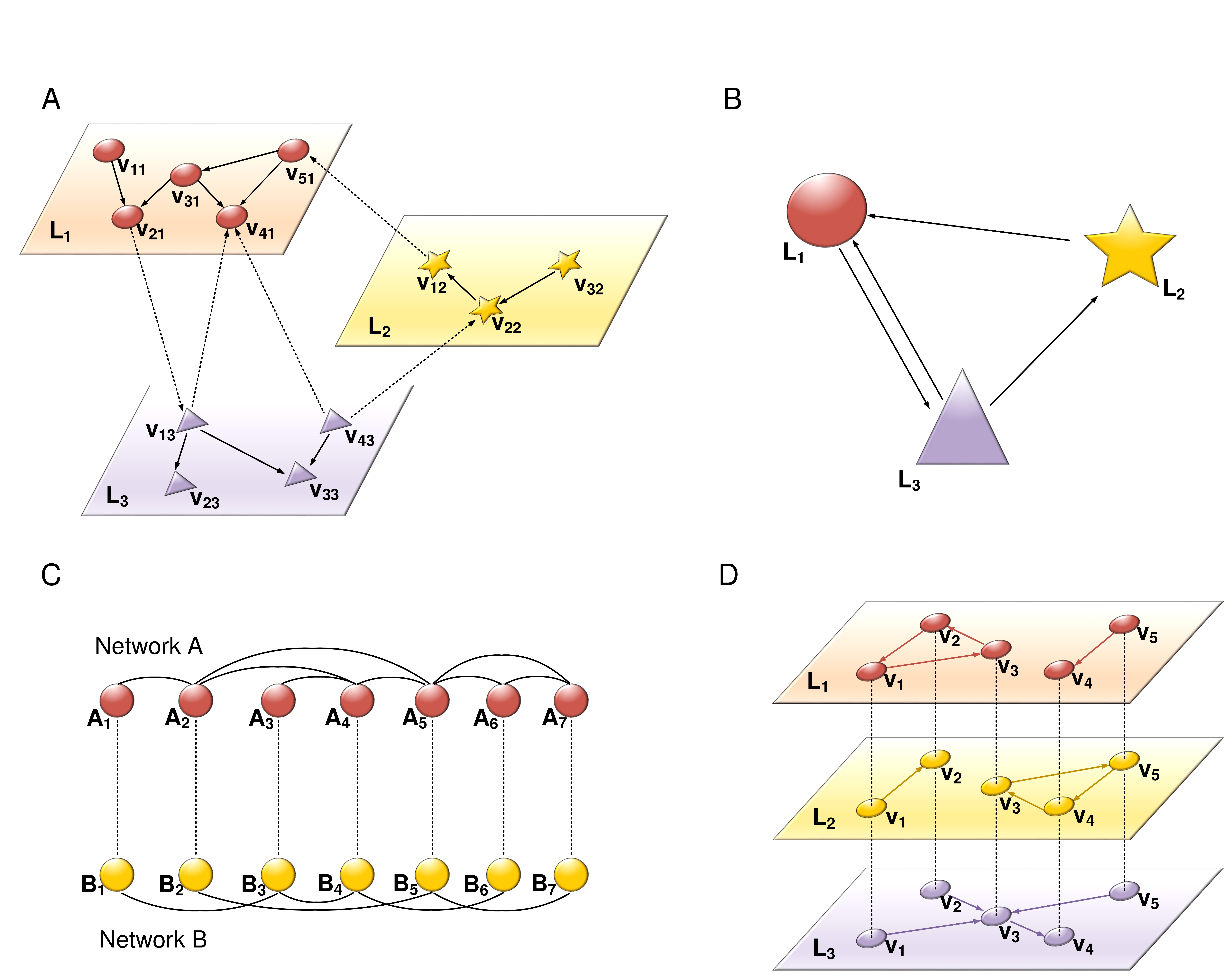}
\caption{\textbf{Examples of multilayer networks.} (\textbf{A}) An example of a multilayer network, where nodes in one layer can connect with nodes in other layers.
The links between nodes in different layers are called interlayer links, and those in the same layer are termed intralayer links. (\textbf{B}) A network of networks formed from (\textbf{A}): nodes represent layers, and the links indicate the interlayer influence.
(\textbf{C}) An example of an interdependent network. It comprises a collection of networks, of which nodes in one network  are interdependent with nodes in other networks.
(\textbf{D}) An example of a multiplex network. It has the same set of nodes in all layers, and its interlayer links (dotted lines) exist only between counterparts in different layers.}
\end{figure*}

\clearpage
\begin{figure*}
\centering
\includegraphics[width=\textwidth]{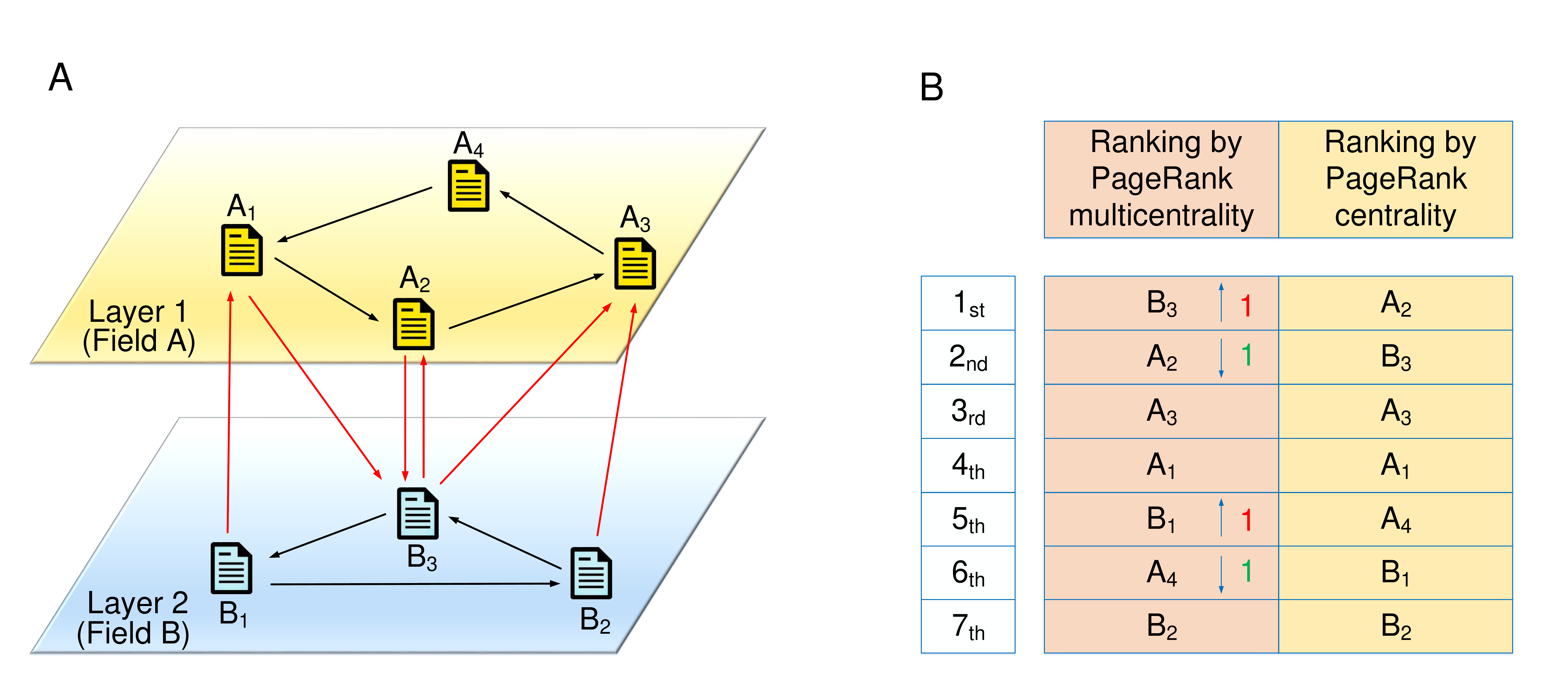}
\caption{\textbf{PageRank multicentrality in a multilayer network compared with PageRank centrality in a single-layer network (constructed by aggregating nodes from multiple layers into a single layer).} A toy example of a multilayer citation network with two layers is provided in (\textbf{A}).
Nodes in layer 1 denote papers in field A, while nodes in layer 2 represent papers in field B. The black links indicate citations between papers in the same field and the red links denote citations between papers in different fields.
The results are shown  in (\textbf{B}).
In global PageRank multicentrality,  the paper $B_3$ is ranked as the most important one.
However, in PageRank centrality, paper $A_2$ has the largest centrality.
Note that the aggregated centrality of nodes in layer 1 (thus the importance) is higher than that in layer 2.
The  propagating multicentrality from layer 2 to layer 1 is greater than that from layer 1 to layer 2.
Since paper $B_3$ has been cited by two papers in layer 1, its multicentrality is magnified and is thus  higher than that in PageRank centrality.}
\label{fig.pagepaper}
\end{figure*}

\clearpage

\begin{figure*}
\centering
\includegraphics[width=\textwidth]{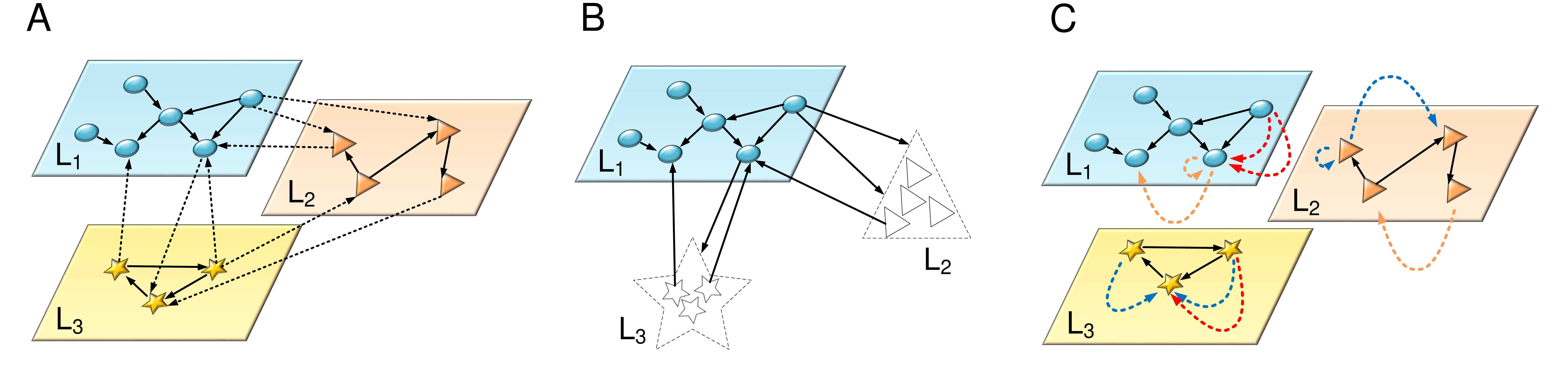}
\caption{ \textbf{A schematic illustrating local multicentrality in a multilayer network.} (\textbf{A}) shows an example of a multilayer network with three layers. (\textbf{B}) shows the local multicentrality graph schematics. When calculating multicentrality of nodes in $L_1$ (layer 1), we set initial scores to nodes in $L_1$ and they attribute scores to neighbors along both intralayer links and interlayer links iteratively. In this process, we regard the other layers ($L_2$ and $L_3$) as ``virtual layers'', where nodes will distribute scores to $L_1$ along the interlayer links once they gather scores from nodes in $L_1$. In the calculation of the multicentrality in other layers, we can adopt an analogous process. The local multicentrality of a node is the composite of the scores it gathers in the steady state. (\textbf{C}) shows the steady state in calculating multicentrality. In each layer, the propagating scores will not leave the layer, and the colored links mean that the scores  are dependent on the multicentrality of the nodes in the corresponding layer. }
\label{fig.local}
\end{figure*}

\clearpage

\begin{figure*}
\centering
\includegraphics[width=\textwidth]{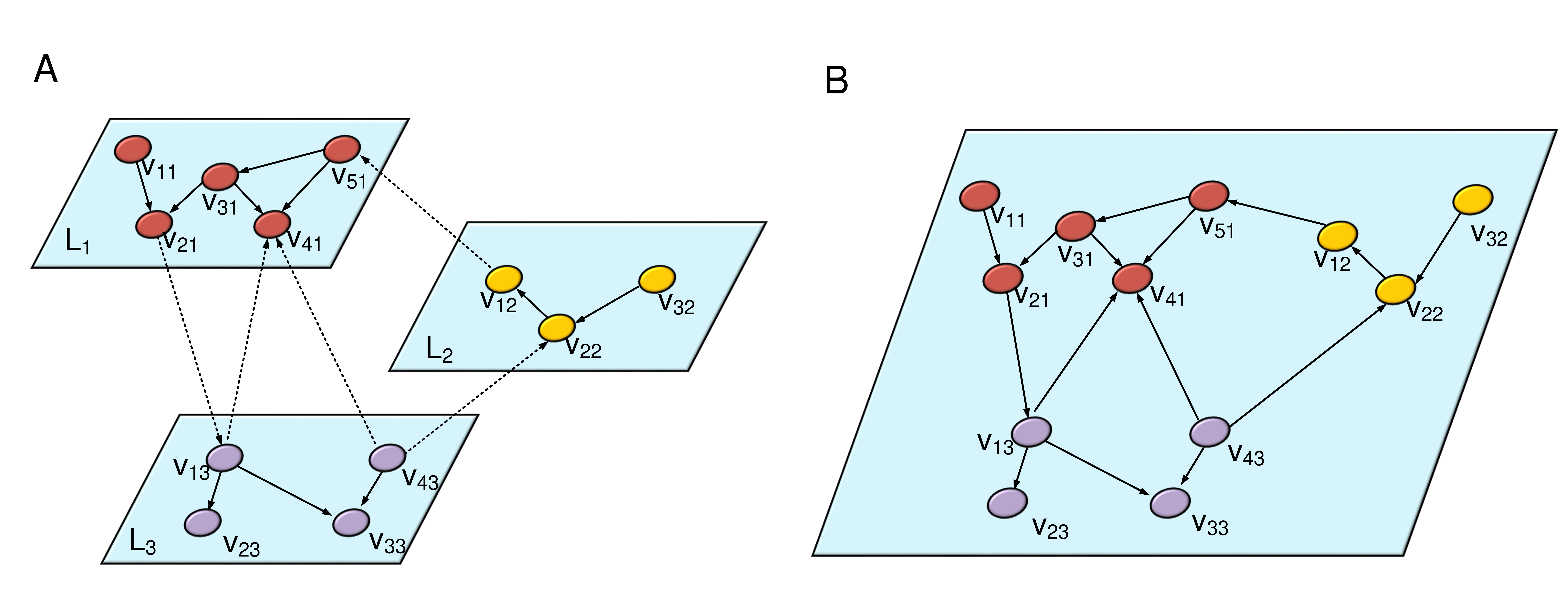}
\caption{\textbf{A schematic illustrating building an aggregated network from a multilayer network.} (\textbf{A}) shows a general multilayer network with three layers. (\textbf{B}) shows the aggregated (single-layer) network built from the multilayer network in (\textbf{A}). In the aggregated network here, all links are regarded as  the same type and all nodes are centrality-homogeneous.}
\label{fig.aggregated}
\end{figure*}

\clearpage

\begin{figure*}
\centering
\includegraphics[width=\textwidth]{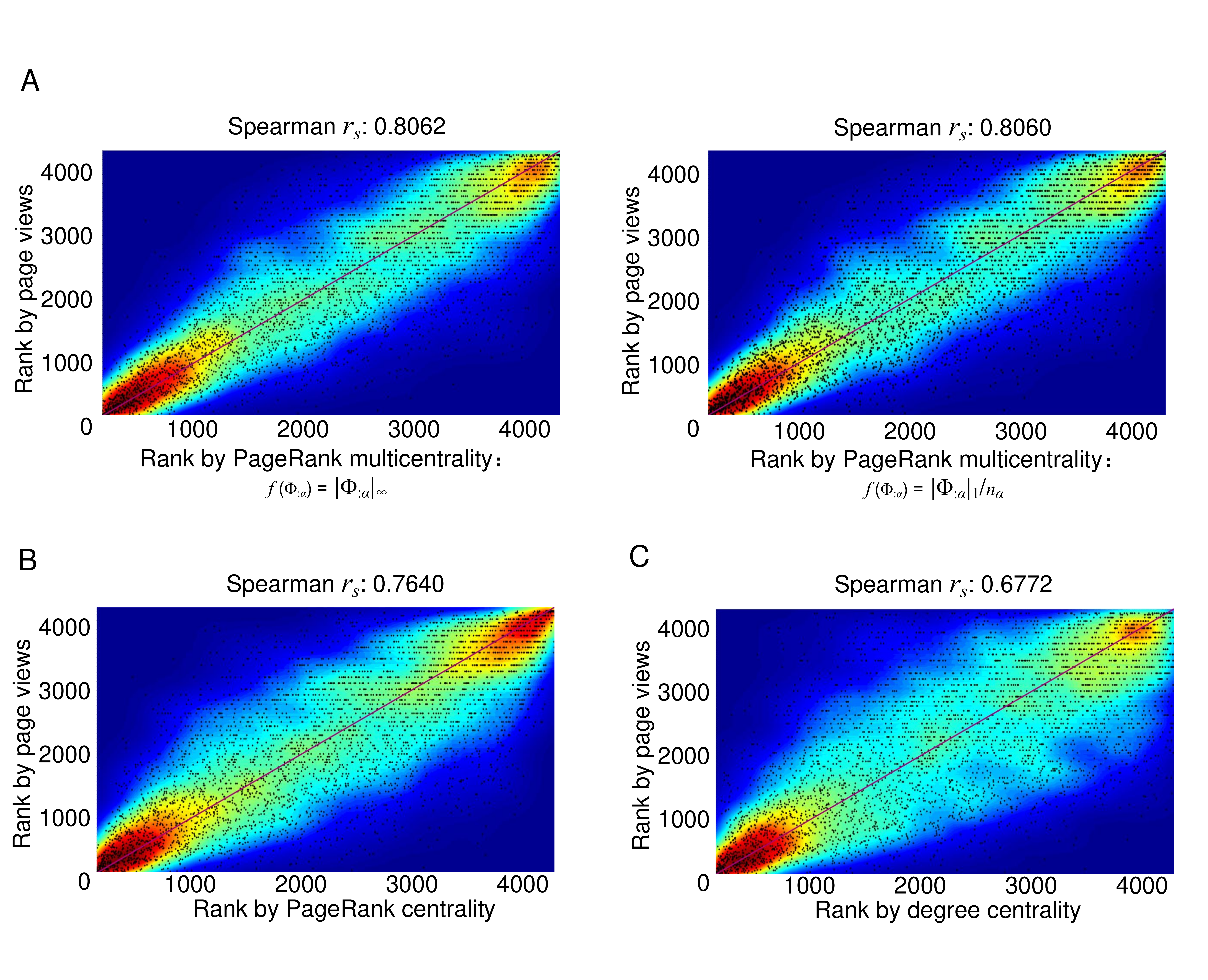}
\caption{\textbf{Comparison between PageRank multicentrality, PageRank centrality and degree centrality.} (\textbf{A}) shows two scatter diagrams of PageRank multicentrality of all nodes under the other two forms of $f(\cdot)$, where the horizontal axis denotes the rank by  multicentrality and the vertical axis represents the rank by PVs in Wikispeedia. The two spearman correlation coefficients between these two rankings are  0.8062 and 0.8060. (\textbf{B}) shows a scatter diagram of PageRank centrality on the single-layer network, and the Spearman correlation coefficient is 0.7640. (\textbf{C}) shows a scatter diagram of the ranks by degree centrality and  by PVs, and the Spearman correlation coefficient is 0.6772.}
\label{fig.Comparison}
\end{figure*}

\clearpage

\begin{figure*}
\centering
\includegraphics[width=11.4 cm]{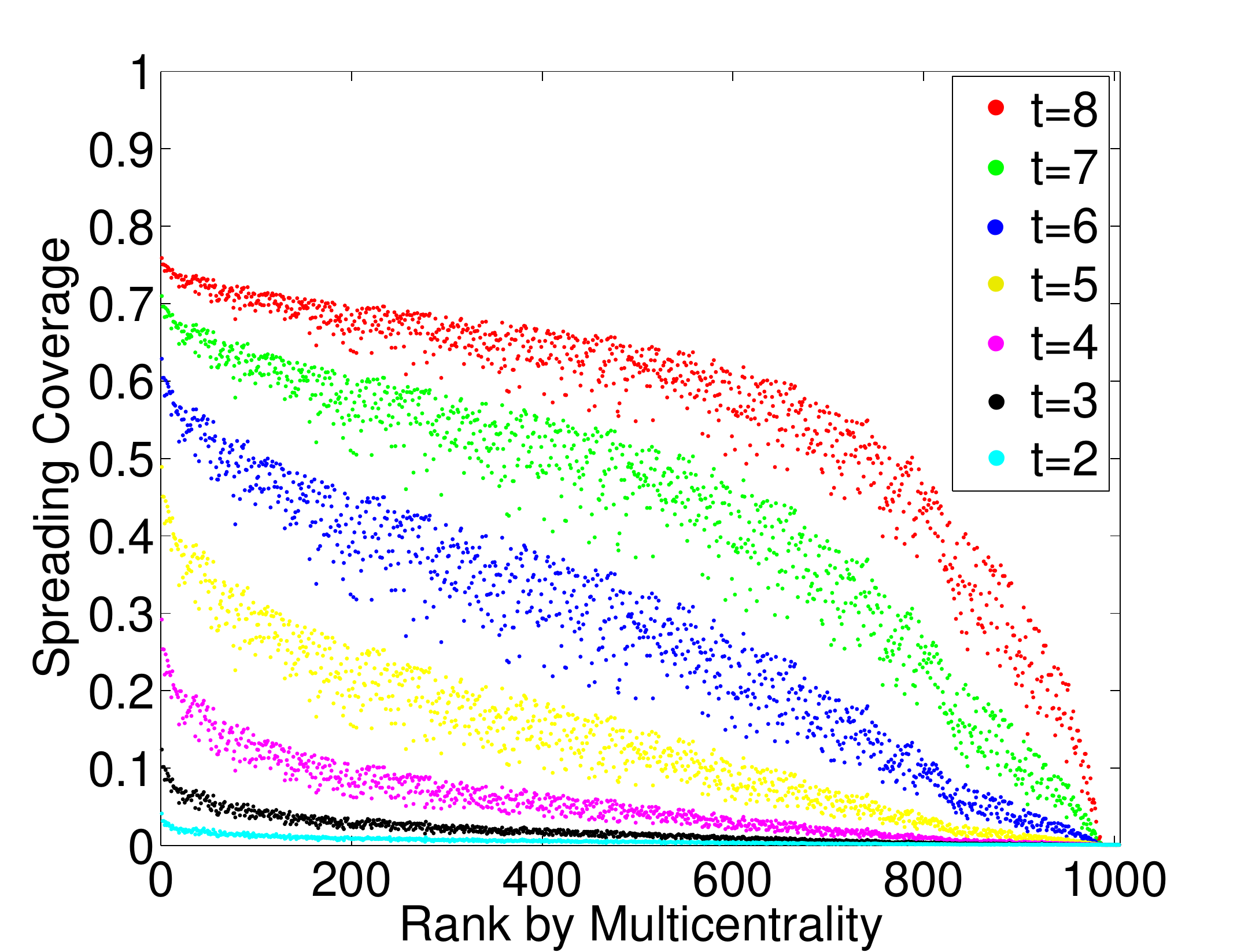}
\caption{\textbf{The relationship between the epidemic spreading process and eigenvector multicentrality.} The scatter diagram  shows the results of an epidemic spreading process and eigenvector multicentrality, where different colors represent the corresponding time steps ($t$). The horizontal axis denotes the rank by eigenvector multicentrality and the vertical axis denotes the spreading coverage. Each dot in the scatter diagram indicates the spreading coverage at each time step, assuming that  the infection source is a certain node. Obviously, the spreading coverage will be high if we choose the node with high multicentrality as the infection source. This is validated by the Spearman correlation coefficient which reaches 0.9771 between the rankings by multicentrality and spreading coverage.}
\label{fig.spreading}
\end{figure*}

\clearpage

\subsection{Supplementary Tables}
\vspace{0pt}
$\ $
\begin{table}[h!]\label{table.wikic1}
\centering
\caption{Wikipedia dataset: comparison between the ranking by global PageRank multicentrality ($f(\Phi_{:\alpha})=\ln(1+N\cdot|\Phi_{:\alpha}|_1/n_\alpha)$) in the multilayer network and the ranking by PageRank centrality in the aggregated network. }
\begin{center}
\begin{tabular}{p{4.5cm}<{\centering} p{5cm}<{\centering} p{4cm}<{\centering} p{2.6cm}<{\centering}}
\hline
Entry & Layer & Global PageRank multicentrality & PageRank centrality\\
\hline
United States & Countries & 1 & 1 (+0) \\
Europe & Geography & 2 & 3 (+1) \\
World War II & History & 3 & 7 (+4) \\
England & Geography & 4 & 11 (+7)\\
English language & Language and literature & 5 & 6 (+1) \\
United Kingdom & Countries & 6 & 4 (-2)\\
Latin & Language and literature & 7 & 8 (+1) \\
France & Countries & 8 & 2 (-6) \\
Animal & Science & 9  & 77 (+68) \\
Time zone & Geography & 10 & 10 (+0)\\
Scientific classification & Science & 11 & 115 (+104) \\
London & Geography & 12 & 28 (+16) \\
Africa & Geography & 13 & 23 (+10) \\
Germany & Countries & 14 & 5 (-9) \\
Currency & Business Studies & 15 & 14 (-1)\\
North America & Geography & 16 & 35 (+19)\\
Christianity &  Religion & 17 & 19 (+2)\\
Earth & Science & 18 & 39 (+21) \\
World War I & History & 19 & 30 (+11)\\
French language & Language and literature & 20 & 24 (+4)\\
List of countries & Citizenship & 21 & 18 (-3)\\
United Nations & Citizenship & 22 & 22 (+0)\\
India & Countries & 23 & 9 (-14) \\
Japan & Countries & 24 & 13 (-11) \\
19th century & History & 25 & 31 (+6)\\
\hline
\end{tabular}
\end{center}
\end{table}

\begin{table}\label{table.wikic2}
\centering
\caption{Wikipedia dataset: comparison between the ranking by global PageRank multicentrality ($f(\Phi_{:\alpha})=|\Phi_{:\alpha}|_\infty$) in the multilayer network and the ranking by PageRank centrality in the aggregated network. }
\begin{center}
\begin{tabular}{p{4.5cm}<{\centering} p{5cm}<{\centering} p{4cm}<{\centering} p{2.6cm}<{\centering}}
\hline
Entry & Layer & Global PageRank multicentrality & PageRank centrality\\
\hline
United States & Countries & 1 & 1 (+0) \\
Europe & Geography & 2 & 3 (+1) \\
English language & Language and literature & 3 & 6  (+3)\\
World War II & History & 4 & 7 (+3)  \\
United Kingdom & Countries & 5  & 4 (-1)\\
France & Countries & 6 & 2 (-4) \\
England & Geography & 7 & 11 (+4)\\
Latin & Language and literature & 8 & 8 (+0) \\
Animal & Science & 9  & 77 (+68) \\
Scientific classification & Science & 10 & 115  (+105) \\
Currency & Business Studies & 11 & 14 (+3)\\
Christianity &  Religion & 12 & 19 (+7)\\
United Nations & Citizenship & 13 & 22 (+9) \\
List of countries & Citizenship & 14 & 18 (+4)\\
Islam & Religion & 15 & 21 (+6) \\
Germany & Countries & 16 & 5 (-11) \\
Agriculture & Everyday life & 17 & 49 (+32)\\
Africa & Geography & 18 & 23 (+5) \\
London & Geography & 19 & 28 (+9) \\
Time zone & Geography & 20 &10 (-10)\\
European Union & Citizenship & 21 & 29 (+8)\\
Football  soccer & Everyday life & 22  & 54 (+32)\\
Earth & Science & 23 & 39 (+16) \\
Carolus Linnaeus & People & 24  & 324 (+300)\\
India & Countries & 25 & 9 (-16) \\
\hline
\end{tabular}
\end{center}
\end{table}

\begin{table}\label{table.wikic3}
\centering
\caption{Wikipedia dataset: comparison between the ranking by global PageRank multicentrality ($f(\Phi_{:\alpha})=|\Phi_{:\alpha}|_1/n_\alpha$) in the multilayer network and the ranking by PageRank centrality in the aggregated network. }
\begin{center}
\begin{tabular}{p{4.5cm}<{\centering} p{5cm}<{\centering} p{4cm}<{\centering} p{2.6cm}<{\centering}}
\hline
Entry & Layer & Global PageRank multicentrality & PageRank centrality\\
\hline
Europe & Geography & 1 & 3 (+2) \\
United States & Countries & 2 & 1 (-1) \\
World War II & History & 3 & 7 (+4) \\
England & Geography & 4 & 11 (+7)\\
English language & Language and literature & 5 & 6 (+1) \\
Latin & Language and literature & 6 & 8 (+2) \\
United Kingdom & Countries & 7 & 4 (-3)\\
Animal & Science & 8 & 77 (+69) \\
Time zone & Geography & 9 &10 (+1)\\
France & Countries & 10 & 2 (-8) \\
Scientific classification & Science & 11  & 115 (+104) \\
London & Geography & 12 & 28  (+16) \\
Africa & Geography & 13  & 23 (+10) \\
Currency & Business Studies & 14  & 14 (+0)\\
North America & Geography & 15 & 35 (+20)\\
Earth & Science & 16 & 39  (+23)\\
Christianity &  Religion & 17 & 19 (+2)\\
World War I & History & 18  & 30 (+12)\\
Germany & Countries & 19  & 5 (-14) \\
List of countries & Citizenship & 20  & 18 (-2)\\
French language & Language and literature & 21 & 24 (+3)\\
Agriculture & Everyday life & 22 & 49 (+27)\\
United Nations & Citizenship & 23 & 22 (-1)\\
19th century & History & 24 & 31 (+7)\\
20th century & History & 25 & 37 (+12)\\
\hline
\end{tabular}
\end{center}
\end{table}

\begin{table}\label{table.wikig1}
\centering
\caption{Wikipedia dataset: comparison between the ranking by global PageRank multicentrality ($f(\Phi_{:\alpha})=\ln(1+N\cdot|\Phi_{:\alpha}|_1/n_\alpha)$) in the multilayer network and the  ranking by page views.}
\begin{center}
\begin{tabular}{p{4cm}<{\centering} p{5cm}<{\centering} p{4cm}<{\centering} p{2.6cm}<{\centering}}
\hline
Entry & Layer & Global PageRank multicentrality &Page views\\
\hline
United States & Countries & 1 & 1 (+0) \\
Europe & Geography & 2 & 2 (+0) \\
World War II & History & 3 & 7 (+4) \\
England & Geography & 4  & 4 (+0)\\
English language & Language and literature & 5 & 16 (+11) \\
United Kingdom & Countries & 6 & 3 (-3)\\
Latin & Language and literature & 7  & 71 (+64) \\
France & Countries & 8 & 13 (+5) \\
Animal & Science & 9 & 10 (+1) \\
Time zone & Geography & 10  &389 (+379)\\
Scientific classification & Science & 11 & 162 (+151) \\
London & Geography & 12 & 38 (+26) \\
Africa & Geography & 13  & 6 (-7) \\
Germany & Countries & 14 & 9  (-5)\\
Currency & Business Studies & 15 & 233  (+218)\\
North America & Geography & 16 & 8  (-8)\\
Christianity &  Religion & 17 & 29  (+12)\\
Earth & Science & 18 & 5  (-13) \\
World War I & History & 19 & 89  (+70)\\
French language & Language and literature & 20 & 152  (+132)\\
List of countries & Citizenship & 21 & 433  (+412)\\
United Nations & Citizenship & 22 & 31  (+9)\\
India & Countries & 23 & 21 (-2)  \\
Japan & Countries & 24 & 30 (+6)  \\
19th century & History & 25 & 94  (+69)\\
\hline
\end{tabular}
\end{center}
\end{table}

\begin{table}\label{table.wikig2}
\centering
\caption{Wikipedia dataset: comparison between the ranking by global PageRank multicentrality ($f(\Phi_{:\alpha})=|\Phi_{:\alpha}|_\infty$) in the multilayer network and the ranking by page views.}
\begin{center}
\begin{tabular}{p{4cm}<{\centering} p{5cm}<{\centering} p{4cm}<{\centering} p{2.6cm}<{\centering}}
\hline
Entry & Layer & Global PageRank multicentrality & Page views\\
\hline
United States & Countries & 1 & 1 (+0) \\
Europe & Geography & 2 & 2 (+0) \\
English language & Language and literature & 3 & 16  (+13)\\
World War II & History & 4 & 7 (+3) \\
United Kingdom & Countries & 5 & 3 (-2)\\
France & Countries & 6 & 13 (+7) \\
England & Geography & 7 & 4 (-3)\\
Latin & Language and literature & 8 & 71  (+63)\\
Animal & Science & 9 & 10  (+1) \\
Scientific classification & Science & 10 & 162  (+152)\\
Currency & Business Studies & 11 & 233  (+222)\\
Christianity &  Religion & 12  & 29 (+17)\\
United Nations & Citizenship & 13  & 31 (+18) \\
List of countries & Citizenship & 14 & 433  (+419)\\
Islam & Religion & 15 & 119 (+104) \\
Germany & Countries & 16 & 9  (-7) \\
Agriculture & Everyday life & 17 & 24  (+7)\\
Africa & Geography & 18 & 6  (-12)\\
London & Geography & 19 & 38 (+19) \\
Time zone & Geography & 20 &389  (+369)\\
European Union & Citizenship & 21 & 43  (+22)\\
Football soccer & Everyday life & 22 & 222  (+200)\\
Earth & Science & 23  & 5 (-18)\\
Carolus Linnaeus & People & 24 & 744 (+720)\\
India & Countries & 25 & 21 (-4) \\
\hline
\end{tabular}
\end{center}
\end{table}

\begin{table}\label{table.wikig3}
\centering
\caption{Wikipedia dataset: comparison between the ranking by global PageRank multicentrality ($f(\Phi_{:\alpha})=|\Phi_{:\alpha}|_1/n_\alpha$) in the multilayer network and the ranking by page views.}
\begin{center}
\begin{tabular}{p{4cm}<{\centering} p{5cm}<{\centering} p{4cm}<{\centering} p{2.6cm}<{\centering}}
\hline
Entry & Layer & Global PageRank multicentrality & Page views\\
\hline
Europe & Geography & 1 & 2 (+1) \\
United States & Countries & 2 & 1 (-1) \\
World War II & History & 3 & 7 (+4) \\
England & Geography & 4 & 4 (+0)\\
English language & Language and literature & 5 & 16 (+11) \\
Latin & Language and literature & 6 & 71 (+65) \\
United Kingdom & Countries & 7 & 3 (-4)\\
Animal & Science & 8 & 10 (+2)  \\
Time zone & Geography & 9 &389 (+380)\\
France & Countries & 10 & 13  (+3)\\
Scientific classification & Science & 11 & 162 (+151) \\
London & Geography & 12  & 38 (+26) \\
Africa & Geography & 13 & 6  (-7) \\
Currency & Business Studies & 14 & 233 (+219)\\
North America & Geography & 15 & 8 (-7)\\
Earth & Science & 16 & 5  (-11)\\
Christianity &  Religion & 17  & 29 (+12)\\
World War I & History & 18 & 89 (+71) \\
Germany & Countries & 19 & 9 (-10) \\
List of countries & Citizenship & 20 & 433 (+413)\\
French language & Language and literature & 21 & 152 (+131)\\
Agriculture & Everyday life & 22 & 24 (+2) \\
United Nations & Citizenship & 23  & 31 (+8)\\
19th century & History & 24  & 94 (+70)\\
20th century & History & 25 & 59 (+34)\\
\hline
\end{tabular}
\end{center}
\end{table}

\begin{table}\label{table.wikig4}
\centering
\caption{Wikipedia dataset: comparison between the ranking by PageRank centrality  in the aggregated network and the  ranking by page views.}
\begin{center}
\begin{tabular}{p{4cm}<{\centering}  p{5cm}<{\centering} p{2.6cm}<{\centering}}
\hline
Entry  &  PageRank centrality & Page views\\
\hline
United States & 1 & 1 (+0) \\
France  & 2 & 13  (+11)\\
Europe & 3 & 2 (-1) \\
United Kingdom & 4 & 3 (-1)\\
Germany & 5 & 9 (+4) \\
English language & 6 & 16 (+10) \\
World War II & 7 & 7 (+0) \\
Latin & 8 & 71 (+63) \\
India & 9 &  21 (+12)\\
Time zone & 10 &389 (+379)\\
England & 11 & 4 (-7)\\
Italy & 12 & 46 (+34)\\
Japan & 13 & 30 (+17)\\
Currency & 14 & 233 (+219)\\
Spain & 15 & 91 (+76)\\
China & 16 & 28 (+12)\\
Russia & 17 & 35 (+18)\\
List of countries & 18 & 433 (+415)\\
Christianity &  19  & 29 (+10)\\
Canada & 20 & 40 (+20)\\
Islam & 21 & 119 (+98)\\
United Nations & 22  & 31 (+9)\\
Africa & 23 & 6  (-17) \\
French language & 24 & 152 (+128)\\
Australia & 25 & 36  (+11) \\
\hline
\end{tabular}
\end{center}
\end{table}

\begin{table}\label{table.wikig5}
\centering
\caption{Wikipedia dataset: comparison between the ranking by degree centrality  in the aggregated network and the  ranking by page views.}
\begin{center}
\begin{tabular}{p{4cm}<{\centering}  p{5cm}<{\centering} p{2.6cm}<{\centering}}
\hline
Entry  &  Degree centrality & Page views\\
\hline
United States & 1 & 1 (+0) \\
United Kingdom & 2 & 3 (+1)\\
Europe & 3 & 2 (-1) \\
France  & 4 & 13  (+9)\\
England & 5 & 4 (-1)\\
Germany & 6 & 9 (+3) \\
World War II & 7 & 7 (+0) \\
English language & 8 & 16 (+8) \\
India & 9 &  21 (+12)\\
London & 10  & 38 (+28) \\
Africa & 11 & 6  (-5) \\
Japan & 12 & 30 (+18)\\
Australia & 13 & 36  (+23) \\
Italy & 14 & 46 (+32)\\
Spain & 15 & 91 (+76)\\
Russia & 16 & 35 (+19)\\
Canada & 17 & 40 (+23)\\
China & 18 & 28 (+10)\\
World War I & 19 & 89 (+70) \\
Asia &  20 & 26 (+6) \\
Scientific classification & 21 & 162 (+141) \\
North America & 22 & 8 (-14) \\
Latin & 23 & 71 (+48) \\
List of countries & 24 & 433 (+409)\\
19th century & 25  & 94 (+69)\\
\hline
\end{tabular}
\end{center}
\end{table}

\begin{table}\label{table.wikil1}
\centering
\caption{Wikipedia dataset: the layer importance of each layer according to global PageRank multicentrality ($f(\Phi_{:\alpha})=\ln(1+N\cdot|\Phi_{:\alpha}|_1/n_\alpha)$) in the  multilayer network.}
\begin{center}
\begin{tabular}{p{5cm}<{\centering} p{3cm}<{\centering} p{4cm}<{\centering} p{2.5cm}<{\centering}}
\hline
Layer & Number of nodes & Layer importance & Ranking\\
\hline
Countries & 222 & $1.1636\times 10^{0}$ & 1 \\
Citizenship & 198 & $7.9775\times 10^{-1}$ & 2 \\
Business Studies & 80 & $7.8469\times 10^{-1}$ & 3\\
Religion & 111 & $7.8034\times 10^{-1}$ & 4  \\
Language and literature & 155 & $7.4571\times 10^{-1}$ & 5  \\
Geography & 758 & $7.0347\times 10^{-1}$ & 6 \\
History & 417 & $6.8889\times 10^{-1}$ & 7  \\
Science & 902& $6.0636\times 10^{-1}$ & 8 \\
Mathematics & 42 & $5.9404\times 10^{-1}$ & 9\\
Art & 36 & $5.8302\times 10^{-1}$ & 10\\
Everyday life & 329 &  $5.5456\times 10^{-1}$ & 11\\
People & 520 & $5.2900\times 10^{-1}$ & 12\\
IT & 66 & $5.2776\times 10^{-1}$ & 13\\
Design and Technology & 202 & $4.8217\times 10^{-1}$ & 14\\
Music & 86 & $4.4818\times 10^{-1}$ & 15\\
\hline
\end{tabular}
\end{center}
\end{table}

\begin{table}\label{table.wikil2}
\centering
\caption{Wikipedia dataset: the layer importance of each layer according to global PageRank multicentrality ($f(\Phi_{:\alpha})=|\Phi_{:\alpha}|_\infty$) in the  multilayer network.}
\begin{center}
\begin{tabular}{p{5cm}<{\centering} p{3cm}<{\centering} p{4cm}<{\centering} p{2.5cm}<{\centering}}
\hline
Layer & Number of nodes & Layer importance & Ranking\\
\hline
Countries & 222 & $6.0460\times 10^{-3}$ & 1 \\
Geography & 758 & $4.8709\times 10^{-3}$ & 2 \\
Language and literature & 155 & $4.2282\times 10^{-3}$ & 3  \\
History & 417 & $4.1889\times 10^{-3}$ & 4  \\
Business Studies & 80 & $3.3748\times 10^{-3}$ & 5\\
Science & 902& $3.3699\times 10^{-3}$ & 6 \\
Religion & 111 & $3.2398\times 10^{-3}$ & 7  \\
Citizenship & 198 & $3.1799\times 10^{-3}$ & 8 \\
Everyday life & 329 &  $2.6821\times 10^{-3}$ & 9\\
People & 520 & $2.3712\times 10^{-3}$ & 10\\
Mathematics & 42 & $2.0417\times 10^{-3}$ & 11\\
Design and Technology & 202 & $1.9605\times 10^{-3}$ & 12\\
Art & 36 & $1.8568\times 10^{-3}$ & 13\\
Music & 86 & $1.7267\times 10^{-3}$ & 14\\
IT & 66 & $1.6635\times 10^{-3}$ & 15\\
\hline
\end{tabular}
\end{center}
\end{table}

\begin{table}\label{table.wikil3}
\centering
\caption{Wikipedia dataset: the layer importance of each layer according to global PageRank multicentrality ($f(\Phi_{:\alpha})=|\Phi_{:\alpha}|_1/n_\alpha$) in the  multilayer network.}
\begin{center}
\begin{tabular}{p{5cm}<{\centering} p{3cm}<{\centering} p{4cm}<{\centering} p{2.5cm}<{\centering}}
\hline
Layer & Number of nodes & Layer importance & Ranking\\
\hline
Countries & 222 & $4.7254\times 10^{-4}$ & 1 \\
Citizenship & 198 & $2.9022\times 10^{-4}$ & 2 \\
Business Studies & 80 & $2.8498\times 10^{-4}$ & 3\\
Religion & 111 & $2.8201\times 10^{-4}$ & 4  \\
Language and literature & 155 & $2.6628\times 10^{-4}$ & 5  \\
Geography & 758 & $2.4802\times 10^{-4}$ & 6 \\
History & 417 & $2.4191\times 10^{-4}$ & 7  \\
Science & 902& $2.0548\times 10^{-4}$ & 8 \\
Mathematics & 42 & $2.0029\times 10^{-4}$ & 9\\
Art & 36 & $1.9728\times 10^{-4}$ & 10\\
Everyday life & 329 &  $1.8681\times 10^{-4}$ & 11\\
People & 520 & $1.7659\times 10^{-4}$ & 12\\
IT & 66 & $1.7371\times 10^{-4}$ & 13\\
Design and Technology & 202 & $1.5855\times 10^{-4}$ & 14\\
Music & 86 & $1.4534\times 10^{-4}$ & 15\\
\hline
\end{tabular}
\end{center}
\end{table}

\begin{table}\label{table.ea1}
\centering
\caption{European airline dataset: comparison of the rankings by PageRank multicentrality ($f(\Phi_{:\alpha})=e^{|\Phi_{:\alpha}|_1/n_\alpha}-1$), PageRank versatility and degree centrality.}
\begin{center}
\begin{tabular}{p{6cm}<{\centering} p{3.7cm}<{\centering} p{2.7cm}<{\centering} p{2.7cm}<{\centering}}
\hline
Airport & PageRank multicentrality & PageRank versatility & Degree   centrality\\
\hline
London Stansted & 1 & 2 (+1) & 1 (+0) \\
London Gatwick & 2 & 4 (+2) & 4 (+2) \\
Flughafen M$\ddot{u}$nchen & 3 & 1 (-2) & 2 (-1)\\
Frankfurt  & 4 & 3 (-1) & 3 (-1)\\
Malpensa International & 5 & 5 (+0) & 9 (+4)\\
Aeropuerto de Madrid & 6 & 7(+1) & 6 (+0)\\
Flughafen D$\ddot{u}$sseldorf International & 7 & 6 (-1) & 8 (+1)\\
Dublin & 8 & 8 (+0) & 5 (-3)\\
Liverpool John Lennon & 9 & 11 (+2) & 14 (+5)\\
London Luton & 10 & 10 (+0) & 10 (+0)\\
Barcelona & 11 & 9 (-2) & 11 (+0)\\
Son Sant Joan & 12 & 12 (+0) & 15 (+3)\\
El Altet & 13 & 13 (+0) & 13 (+0)\\
Charles De Gaulle & 14 & 15 (+1) & 17 (+3)\\
Bristol International & 15 & 18 (+3) & 23 (+8)\\
Orio al Serio International & 16 & 14 (-2) & 7 (-9)\\
Edinburgh & 17 & 16 (-1) & 18 (+1)\\
Malaga & 18 & 17 (-1) & 16 (-2)\\
Brussels South Charleroi  & 19 & 19 (+0) & 12 (-7)\\
Flughafen Berlin Brandenburg & 20 & 20 (+0) & 20 (+0)\\
Manchester International & 21 & 21 (+0) & 36 (+15)\\
Venice Marco Polo International & 22 & 22 (+0) & 26 (+4)\\
Rome Ciampino & 23 & 23 (+0) & 19 (-4)\\
Pisa International & 24 & 24 (+0) & 22 (-2)\\
Bologna Guglielmo Marconi & 25 & 26 (+1) & 28 (+3)\\
\hline
\end{tabular}
\end{center}
\end{table}

\begin{table}\label{table.ea2}
\centering
\caption{European airline dataset: comparison of the rankings by PageRank multicentrality ($f(\Phi_{:\alpha})=|\Phi_{:\alpha}|_1/n_\alpha$), PageRank versatility and degree centrality.}
\begin{center}
\begin{tabular}{p{6cm}<{\centering} p{3.7cm}<{\centering} p{2.7cm}<{\centering} p{2.7cm}<{\centering}}
\hline
Airport & PageRank multicentrality & PageRank versatility & Degree  centrality\\
\hline
London Stansted & 1 & 2 (+1) & 1 (+0) \\
London Gatwick & 2 & 4 (+2) & 4 (+2) \\
Flughafen M$\ddot{u}$nchen & 3 & 1 (-2) & 2 (-1)\\
Frankfurt  & 4 & 3 (-1) & 3 (-1)\\
Malpensa International & 5 & 5 (+0) & 9 (+4)\\
Aeropuerto de Madrid & 6 & 7 (+1) & 6 (+0)\\
Dublin & 7 & 8 (+1) & 5 (-2)\\
Liverpool John Lennon & 8 & 11 (+3) & 14 (+6)\\
Flughafen D$\ddot{u}$sseldorf International & 9 & 6 (-3) & 8 (-1)\\
London Luton & 10 & 10 (+0) & 10 (+0)\\
Barcelona & 11 & 9 (-2) & 11 (+0)\\
Son Sant Joan & 12 & 12 (+0) & 15 (+3)\\
El Altet & 13 & 13 (+0) & 13 (+0)\\
Orio al Serio International & 14 & 14 (+0) & 7 (-7)\\
Charles De Gaulle & 15 & 15 (+0) & 17 (+2)\\
Bristol International & 16 & 18 (+2) & 23 (+7)\\
Edinburgh & 17 & 16 (-1) & 18 (+1)\\
Malaga & 18 & 17 (-1) & 16 (-2)\\
Brussels South Charleroi  & 19 & 19 (+0) & 12 (-7)\\
Flughafen Berlin Brandenburg & 20 & 20 (+0) & 20 (+0)\\
Rome Ciampino & 21 & 23 (+2) & 19 (-2)\\
Manchester International & 22 & 21 (-1) & 36 (+14)\\
Venice Marco Polo International & 23 & 22 (-1) & 26 (+3)\\
Pisa International & 24 & 24 (+0) & 22 (-2)\\
Bologna Guglielmo Marconi & 25 & 26 (+1) & 28 (+3)\\
\hline
\end{tabular}
\end{center}
\end{table}

\begin{table}\label{table.ea3}
\centering
\caption{European airline dataset: comparison of the rankings by PageRank multicentrality ($f(\Phi_{:\alpha})=\ln(1+N\cdot|\Phi_{:\alpha}|_1/n_\alpha)$), PageRank versatility and degree centrality.}
\begin{center}
\begin{tabular}{p{6cm}<{\centering} p{3.7cm}<{\centering} p{2.7cm}<{\centering} p{2.7cm}<{\centering}}
\hline
Airport & PageRank multicentrality & PageRank versatility & Degree  centrality\\
\hline
London Stansted & 1 & 2 (+1) & 1 (+0) \\
London Gatwick & 2 & 4 (+2) & 4 (+2) \\
Flughafen M$\ddot{u}$nchen & 3 & 1 (-2) & 2 (-1)\\
Frankfurt  & 4 & 3 (-1) & 3 (-1)\\
Malpensa International & 5 & 5 (+0) & 9 (+4)\\
Aeropuerto de Madrid & 6 & 7 (+1) & 6 (+0)\\
Dublin & 7 & 8 (+1) & 5 (-2)\\
Liverpool John Lennon & 8 & 11 (+3) & 14 (+6)\\
London Luton & 9 & 10 (+1) & 10 (+1)\\
Barcelona & 10 & 9 (-1) & 11 (+1)\\
Flughafen D$\ddot{u}$sseldorf International & 11 & 6 (-5) & 8 (-3)\\
El Altet & 12 & 13 (+1) & 13 (+1)\\
Son Sant Joan & 13 & 12 (-1) & 15 (+2)\\
Orio al Serio International & 14 & 14 (+0) & 7 (-7)\\
Charles De Gaulle & 15 & 15 (+0) & 17 (+2)\\
Bristol International & 16 & 18 (+2) & 23 (+7)\\
Edinburgh & 17 & 16 (-1) & 18 (+1)\\
Malaga & 18 & 17 (-1) & 16 (-2)\\
Brussels South Charleroi  & 19 & 19 (+0) & 12 (-7)\\
Flughafen Berlin Brandenburg & 20 & 20 (+0) & 20 (+0)\\
Rome Ciampino & 21 & 23 (+2) & 19 (-2)\\
Manchester International & 22 & 21 (-1) & 36 (+14)\\
Venice Marco Polo International & 23 & 22 (-1) & 26 (+3)\\
Pisa International & 24 & 24 (+0) & 22 (-2)\\
Girona-Costa Brava & 25 & 27 (+2) & 21 (-4)\\
\hline
\end{tabular}
\end{center}
\end{table}

\begin{table}\label{table.aa}
\centering
\caption{American airline dataset: comparison of the rankings by PageRank multicentrality ($f(\Phi_{:\alpha})=|\Phi_{:\alpha}|_1/n_\alpha$), PageRank versatility and degree centrality.}
\begin{center}
\begin{tabular}{p{6cm}<{\centering} p{3.7cm}<{\centering} p{2.7cm}<{\centering} p{2.7cm}<{\centering}}
\hline
Airport & PageRank multicentrality & PageRank versatility & Degree  centrality\\
\hline
Atlanta  & 1 & 1 (+0) & 1 (+0) \\
O'Hare  & 2 & 2 (+0) & 2 (+0) \\
Denver  & 3 & 3 (+0) & 4 (+1)\\
Detroit Metropolitan  & 4 & 4 (+0) & 5 (+1)\\
Dallas-Fort Worth  & 5 & 5 (+0) & 3 (-2)\\
Phoenix Sky Harbor & 6 & 6 (+0) & 11 (+5)\\
George Bush & 7 & 8 (+1) & 6 (-1)\\
Salt Lake City  & 8 & 7 (-1) & 8 (+0)\\
Minneapolis-Saint Paul   & 9 & 9 (+0) & 7 (-2)\\
Los Angeles  & 10 & 10 (+0) & 13 (+3)\\
Las Vegas McCarran  & 11 & 11 (+0) & 10 (-1)\\
Newark Liberty & 12 & 12 (+0) & 12 (+0)\\
Cincinnati & 13 & 13 (+0) & 9 (-4)\\
Orlando & 14 & 14 (+0) & 15 (+1)\\
John Kennedy & 15 & 15 (+0) & 21 (+6)\\
Charlotte/Douglas & 16 & 16 (+0) & 16 (+0)\\
Manpheus & 17 & 17 (+0) & 14 (-3)\\
Boston Logan & 18 & 18 (+0) & 24 (+6)\\
San Francisco  & 19 & 19 (+0) & 20 (+1)\\
Cleveland Hopkins & 20 & 20 (+0) & 18 (-2)\\
Washington Dulles & 21 & 21 (+0) & 19 (-2)\\
Philadelphia & 22 & 22 (+0) & 26 (+4)\\
LaGuardia & 23 & 23 (+0) & 30 (+7)\\
Seattle-Tacoma & 24 & 24 (+0) & 25 (+1)\\
Baltimore-Washington  & 25 & 25 (+0) & 17 (-8)\\
\hline
\end{tabular}
\end{center}
\end{table}

\appendix

\clearpage
\section*{Acknowledgements.}
We thank Manlio De Domenico for sharing the processed dataset of the European airline network and the details for calculating the versatility. This work was  supported by the National Natural Science Foundation of China (Grant No. 61731004) and the Zhejiang Natural Science Foundation (Grant No. LR16F020001).  This work was supported in part by the U.S. Army Research Office under Grant W911NF-16-1-0448 and DTRA under Grant HDTRA1-13-1-0029.

\section*{Author contributions.}
S.H., J.C., J.Z., and H.V.P. conceived and designed the project.
M.W. conducted all the theoretical analysis and empirical calculations.
All authors analyzed the results.
M.W., S.H., J.Z., and J.C. wrote the manuscript.
Y.-Y.L., J.Z., H.V.P., and Y.S. edited the manuscript.

\subsection*{Competing Interests.}
The authors declare no competing financial interests.

\clearpage
\section{References}
\bibliography{cited}

\beginsupplement

\clearpage

\setcounter{page}{1}

\end{document}